\documentclass[12pt,twoside,a4paper]{article}
\pdfoutput=1
\usepackage{graphicx}
\usepackage{url}
\usepackage{color}
\usepackage{cite}

\begin{document}
\thispagestyle{empty} 

\title{
\vskip-3cm
{\baselineskip14pt
\centerline{\normalsize DESY 17-040 \hfill ISSN 0418--9833}
\centerline{\normalsize MITP/17--016 \hfill} 
\centerline{\normalsize March 2017 \hfill}} 
\vskip1.5cm
\boldmath
{\bf Study of heavy meson production 
\\
in p-Pb collisions at $\sqrt{S}$=5.02 TeV in the}
\\
{\bf general-mass variable-flavour-number scheme}
\unboldmath
\author{
G.~Kramer$^1$, 
and H.~Spiesberger$^2$
\vspace{2mm} \\
\normalsize{
  $^1$ II. Institut f\"ur Theoretische
  Physik, Universit\"at Hamburg,
}\\ 
\normalsize{
  Luruper Chaussee 149, D-22761 Hamburg, Germany
} \vspace{2mm}
\\
  {\normalsize 
  $^2$ PRISMA Cluster of Excellence, Institut f\"ur Physik,}
  \\
  {\normalsize 
  Johannes Gutenberg-Universit\"at, 55099 Mainz, Germany,}
  \\
  {\normalsize 
  and Centre for Theoretical and Mathematical Physics and 
  Department of Physics,}
  \\
  {\normalsize 
  University of Cape Town, Rondebosch 7700, South Africa}
\vspace{2mm} 
\\
}}

%\date{\today}
\maketitle
%%%%%%%%%%%%%%%%%%%%%%%%%%%%%%%%%%%%%%%%%%%%%%%%%%%%%%%%%%%%%%%%%%%%%%%
\begin{abstract}
\medskip
\noindent
We study inclusive charm and bottom production, for both $D$ 
and $B$ mesons, in p-Pb collisions at the LHC. Numerical results 
for $p_T$-differential production cross sections are obtained at 
next-to-leading-order in the general-mass variable-flavor-number 
scheme. We compare our results with recent data from ALICE, LHCb 
and CMS at a center-of-mass energy of 5 TeV and find good 
agreement. A comparison with p-p cross sections does not 
reveal the presence of nuclear initial-state interaction 
effects that could be expected to become visible as deviations 
of the ratio of p-Pb and p-p cross sections from one.  
\end{abstract}

\clearpage
%%%%%%%%%%%%%%%%%%%%%%%%%%%%%%%%%%%%%%%%%%%%%%%%%%%%%%%%%%%%%%%%%%%%%%%
%%%%%%%%%%%%%%%%%%%%%%%%%%%%%%%%%%%%%%%%%%%%%%%%%%%%%%%%%%%%%%%%%%%%%%%

%**********************************************************************
\section{Introduction}

The study of heavy-quark (charm or bottom) production in p-p 
collisions at LHC energies is a useful test of perturbative 
Quantum Chromodynamics (QCD) since the heavy quark mass provides 
a hard scale that allows calculations within perturbation theory. 
The QCD calculations are based on the factorization approach, in 
which cross sections are calculated as a convolution of three 
terms: the parton distribution functions (PDF) of the incoming 
protons, the partonic hard scattering cross sections computed as 
a perturbative series in the strong interaction coupling constant, 
and the fragmentation functions (FF), describing the relative 
production yield and momentum distribution for a given heavy 
hadron ($D$ or $B$ meson) in a parton. Corresponding recent 
calculations at the perturbative level at next-to-leading order 
(NLO) with next-to-leading-log resummation (FONLL) 
\cite{Cacciari:1998it, Cacciari:2003zu} or in the framework 
of the general-mass-variable-flavour-number scheme (GM-VFNS) 
\cite{Kniehl:2004fy,Kniehl:2005mk} have provided good 
descriptions for bottom meson production in $\bar{\rm p}$-p
collisions at $\sqrt{S} = 1.96$ TeV at the FNAL Tevatron Collider 
\cite{Kniehl:2008zza,Acosta:2004yw,Abulencia:2006ps} and in p-p 
collisions at $\sqrt{S} = 7$ TeV at the CERN Large Hadron Collider 
(LHC) by the CMS, ATLAS and the LHCb collaborations 
\cite{Kniehl:2011bk,Khachatryan:2011mk,Chatrchyan:2011pw, 
Chatrchyan:2011vh,ATLAS:2013cia,Aaij:2012jd}. The production 
cross section of charmed hadrons ($D$ mesons) at the Tevatron 
\cite{Acosta:2003ax} or of the ATLAS collaboration at the LHC 
\cite{Aad:2015zix} is also reasonably well described within 
theoretical and experimental uncertainties \cite{Kniehl:2005ej, 
Kniehl:2012ti}.

The GM-VFNS is essentially the conventional NLO parton-model 
approach, supplemented with finite-mass effects, intended to 
improve the description at small transverse momentum $p_T$. 
The original GM-VFNS prescription \cite{Kniehl:2004fy,Kniehl:2005mk, 
Kniehl:2012ti} is, however, not suitable for calculations of the 
cross section $d\sigma/dp_T$ for heavy-quark hadron production 
at very small transverse momentum $p_T$. This is due to the 
specific choice of scale parameters for initial-state ($\mu_I$)
and final-state ($\mu_F$) factorization. The original prescription 
was to set $ \mu_I = \mu_F = \sqrt{m_Q^2 + p_T^2}$, where $m_Q$ 
is the mass of the heavy quark, charm or bottom. At $p_T = 0$, 
the scale parameters approach $\mu_I = \mu_F = m_Q$, and at this 
point the heavy quark PDFs are put to zero by construction 
in almost all available PDF parametrizations. Therefore the 
transition to the fixed-flavour-number-scheme (FFNS), which is 
the appropriate scheme for calculating $d\sigma/dp_T$ at rather 
small $p_T$, is not reached for $p_T > 0$, since the heavy quark 
PDF in the proton decouples at $p_T = 0$, and not for finite 
$p_T > 0$. 

A smooth transition to the FFNS at finite $p_T$ can be achieved 
by exploiting the freedom to choose the factorization scale. 
In Refs.\ \cite {Kniehl:2015fla,Kramer:2015wda} we have studied 
the prescription to fix the initial-state factorization scale 
at $\mu = 0.5\sqrt{m_Q^2 + p_T^2}$ instead of $\mu = \sqrt{m_Q^2 
+ p_T^2}$. For simplicity we have chosen the scales for initial 
and final state factorization equal to each other, $\mu_I = 
\mu_F$. With this scale choice we could achieve a reasonably 
good description of the data for $B$ meson production 
down to $p_T = 0$ for the CDF data \cite{Acosta:2004yw} in 
$\bar{\rm p}$-p collisions at the Tevatron and of the LHCb data 
\cite{Aaij:2012jd} for p-p collisions at the LHC in the forward 
rapidity region at $\sqrt{S} = 7$ TeV. A comparison of data for 
all $D$ meson states $D^0$, $D^+$, $D^{*+}$ and $D_s^+$ measured 
by the LHCb collaboration at $\sqrt{S} = 5,\, 7$ and $13$ TeV 
with predictions from the GM-VFNS scheme with the original scale 
choice for $p_T > 3$ GeV can be found in \cite{Aaij:2016jht, 
Aaij:2013mga, Aaij:2015bpa}.

The LHC Collaborations have also measured cross sections for 
heavy-quark production in p-Pb and Pb-Pb collisions. The 
ALICE collaboration, e.g., have performed detailed studies 
of the $p_T$-differential and rapidity-differential cross 
sections $d\sigma/dp_T$ and $d\sigma/dy$ for $D$-meson 
production in p-Pb collisions at $\sqrt{S} = 5.02$ TeV 
\cite{Abelev:2014hha,Adam:2016ich}, also for small $p_T$, 
as well as in Pb-Pb collisions at $\sqrt{S} = 2.76$ TeV 
\cite{ALICE:2012ab}. Collisions with two heavy nuclei are 
of particular interest for studies of the Quark-Gluon Plasma 
(QGP), a high-density colour-deconfined medium. On the other 
hand, data from p-Pb collisions can be used to determine the 
nuclear modification factor $R_{\rm pPb}$, i.e., the ratio of 
p-Pb cross sections relative to the corresponding p-p 
cross sections scaled by the mass number of the Pb nucleus 
($A = 208$). Data are in particular interesting at small 
$p_T$ where one expects the largest deviation from $R_{\rm pPb} 
= 1$. The value of $R_{\rm pPb}$ is of interest for several 
reasons. First large deviations from one, in particular for 
larger $p_T$, would indicate the presence of initial-state 
interaction effects which are needed to obtain a reliable 
interpretation of corresponding Pb-Pb collision data. Second, 
the value of $R_{\rm pPb}$ is of interest by itself and could 
help to obtain information on the nuclear PDFs, which are 
modified compared to the proton PDFs in bound nucleons 
depending on the parton fractional momentum $x$ and the 
atomic mass number $A$.

Ideally, measurements of the cross sections to determine the 
nuclear modification factor $R_{\rm pPb}$ should be done at the 
same center-of-mass energy $\sqrt{S}$. Unfortunately, this is 
not the case; data for p-p and p-Pb collisions at the same 
$\sqrt{S}$ are not available. Instead, the reference p-p cross 
section at $\sqrt{S} = 5.02$ TeV was obtained from data at 
$\sqrt{S} = 7$ TeV \cite{ALICE:2011aa} by scaling the energy
based on predictions from perturbative QCD. The scaling factor 
was determined for each $D$-meson species separately from 
the FONLL calculations \cite{Cacciari:2012ny}. In case of $B$ 
meson production in p-Pb collisions at $\sqrt{S} = 5.02$ TeV, 
measured by the CMS collaboration \cite{Khachatryan:2015uja}, 
the reference cross section $d\sigma/dp_T$ for p-p collisions was 
directly taken from the FONNL calculations at $\sqrt{S} = 5.02$ 
TeV \cite{Cacciari:2012ny} without any extrapolation from their 
data at larger $\sqrt{S}$.

Due to the interest in the nuclear modification factor 
$R_{\rm pPb}$ for heavy quark hadron production, in particular 
as we expect to obtain important information about initial-state 
interaction effects in Pb-Pb collisions, it is desirable to 
study $R_{\rm pPb}$ also within other factorization schemes. 
This is the purpose of the present work in which we provide 
results from calculations of p-p cross sections $d\sigma/dp_T$ 
for $D$ and $B$ meson production at $\sqrt{S} = 5.02$ TeV in 
the framework of the GM-VFNS. We compare our results with data 
for the production of various $D$ meson species at $\sqrt{S} 
= 5.02$ TeV measured by the ALICE \cite{Abelev:2014hha,Adam:2016ich} 
and LHCb collaborations \cite{Aaij:2017gcy} and for $B$ meson 
production at $\sqrt{S} = 5.02$ TeV measured by the CMS 
collaboration \cite{Khachatryan:2015uja}. Using our results for 
the $p_T$-differential cross sections, we also study the nuclear 
modification factor $R_{\rm pPb}$.

The outline of our work is as follows. In the next section, 
Sect.~2, we give the details of the calculations for $D$ mesons 
with the kinematic constraints of the ALICE and LHCb experiments. 
Section~3 contains our results for $B$ meson production at 
$\sqrt{S} = 5.02$ TeV and a comparison with the CMS data. 
Section~4 is reserved for a discussion of the results.

%%%%%%%%%%%%%%%%%%%%%%%%%%%%%%%%%%%%%%%%%%%%%%%%%%%%%%%%%%%%%%%%%%%%%%%
\section{{\boldmath $D$} meson production in p-p and p-Pb collisions}

The theoretical background and explicit analytic results of the 
GM-VFNS approach were previously presented in detail, see Refs.\  
\cite{Kniehl:2004fy,Kniehl:2005mk} and the references cited 
therein. Here we only describe the input needed for the present 
numerical analysis.

Throughout this paper, we use the PDF set CTEQ14 
\cite{Dulat:2015mca} as implemented in the program library 
LHAPDF \cite{Buckley:2014ana}. The fragmentation functions 
determined in Ref.\ \cite{Kneesch:2007ey} for $D^0$, $D^+$ 
and $D^{*+}$ mesons and in Ref.\ \cite{Kniehl:2006mw} for 
the $D_s^+$ meson were used. These FFs always refer to the 
average of charge-conjugated states. The data from ALICE 
and CMS are understood as averaged cross sections as well, 
$(\sigma(D) + \sigma(\overline{D}))/2$ and $(\sigma(B) + 
\sigma(\overline{B}))/2$, while the LHCb collaboration 
decided to present their data as the sum of charge-conjugated 
states. 

Originally, the default value for the scale parameters for 
renormalization and factorization were set by the transverse 
mass $m_T = \sqrt{m_Q^2 + p_T^2}$. By convention, variations 
around a default value by factors of two up and down were 
considered to obtain an estimate of unknown higher-order 
perturbative contributions and, thereby, assign a theoretical 
uncertainty to numerical results. We introduce the dimensionless 
parameters $\xi_i$ $(i = R, I, F)$ and set $\mu_i = \xi_i m_T$. 
Independent variations of the $\xi_i$ between $1/2$ and 2 are 
restricted by keeping any ratio of the $\xi_i$'s smaller than 2. 
We shall denote this choice of scales as the {\em original}€ 
prescription. 

As already mentioned, this {\em original} scale choice does 
not provide a smooth transition to the FFNS at small $p_T$. 
To achieve this we change the factorization scales to $\mu_I 
= \mu_F = \xi_0 \sqrt{4m_Q^2 + p_T^2}$ with $\xi_0 = 0.49$. 
A similar choice with $\xi_0 = 0.5$ was used in a recent study 
of charm meson production \cite{Moch:2016pok}. In Ref.\ 
\cite{Benzke:2017yjn}, using also $\xi_0 = 0.49$, good 
agreement was found with p-p collision data from the LHCb 
experiments \cite{Aaij:2016jht,Aaij:2013mga,Aaij:2015bpa} 
for $p_T$ values down to $p_T = 0$. The choice of 
$\sqrt{4m_c^2 + p_T^2}$ in place of 
the transverse mass $m_T = \sqrt{m_c^2 + p_T^2}$ is motivated 
by the fact that the kinematic threshold for heavy-quark 
production is at $2 m_c$ in the FFNS approach. With the 
additional factor $\xi_0 = 0.49$ we can ensure that $\mu = 
m_Q$ is reached already slightly above $p_T = 0$. For $m_Q 
= m_c = 1.3$ GeV one has $\mu = m_Q$ at $p_T = 0.528$ GeV. 
We choose this value of $m_c$ to be consistent with the 
value used in the PDF set CTEQ14 from Ref.\ \cite{Dulat:2015mca}; 
otherwise a smooth decoupling of the charm content of the 
proton PDF is not achieved. In our earlier calculations for 
larger values of $p_T$ \cite{Kniehl:2012ti} we had adopted 
$m_c = 1.5$ GeV instead. We determine error bands for 
theoretical uncertainties from variations of the 
renormalization scale only, i.e., by varying $\xi_R$ 
between 1/2 and 2. We have to leave the factorization 
scales unchanged since otherwise the proper transition 
to the FFNS would be lost. This setting of scales will be 
called the {\em modified} scale€ in the following. 

%%%%%%%%%%%%%%%%%%%%%%%%%%%%%%%%%
\begin{figure*}[b!]
\begin{center}
\includegraphics[width=7.5cm]{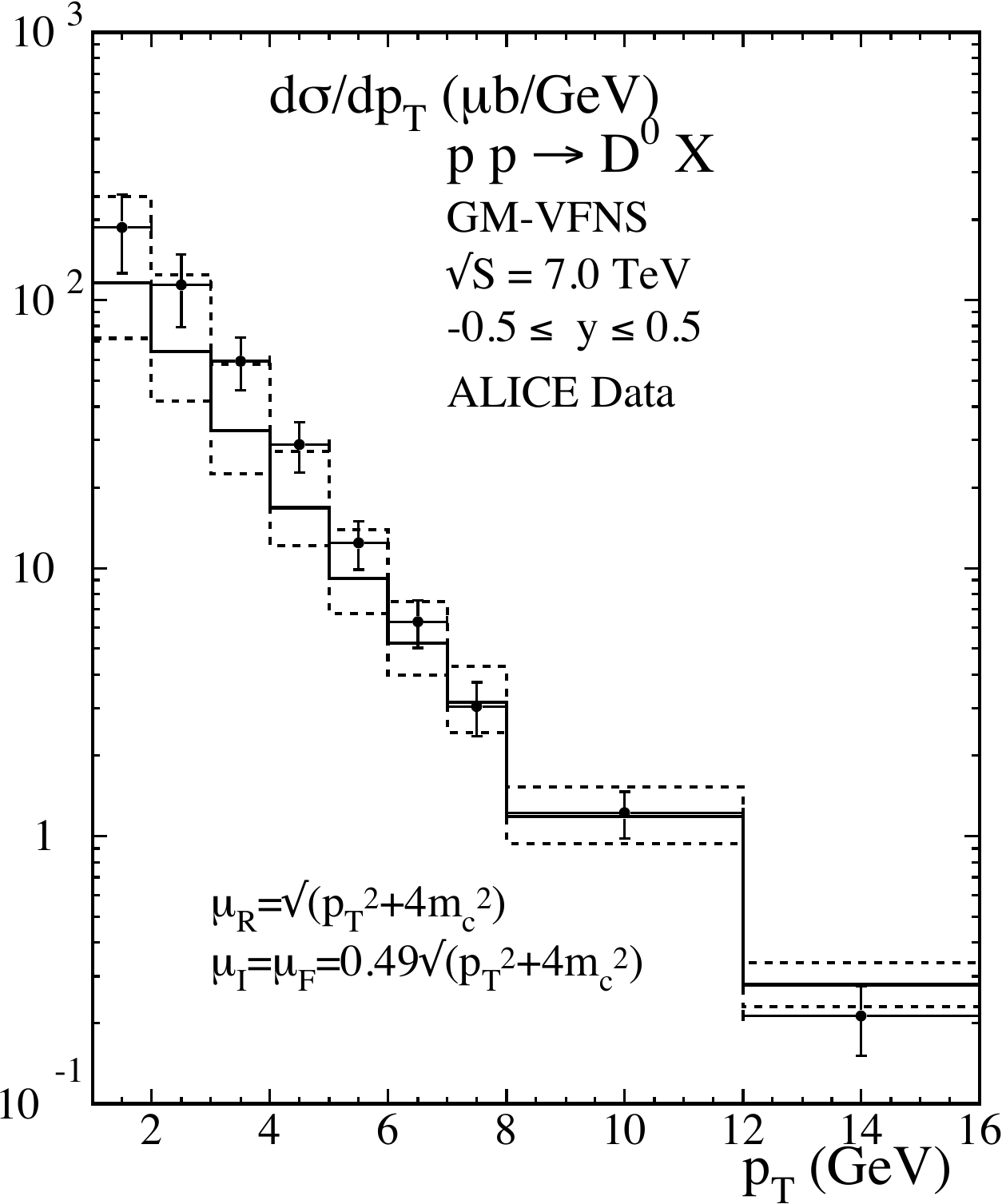}
\raisebox{-1.3mm}{
\includegraphics[width=7.5cm]{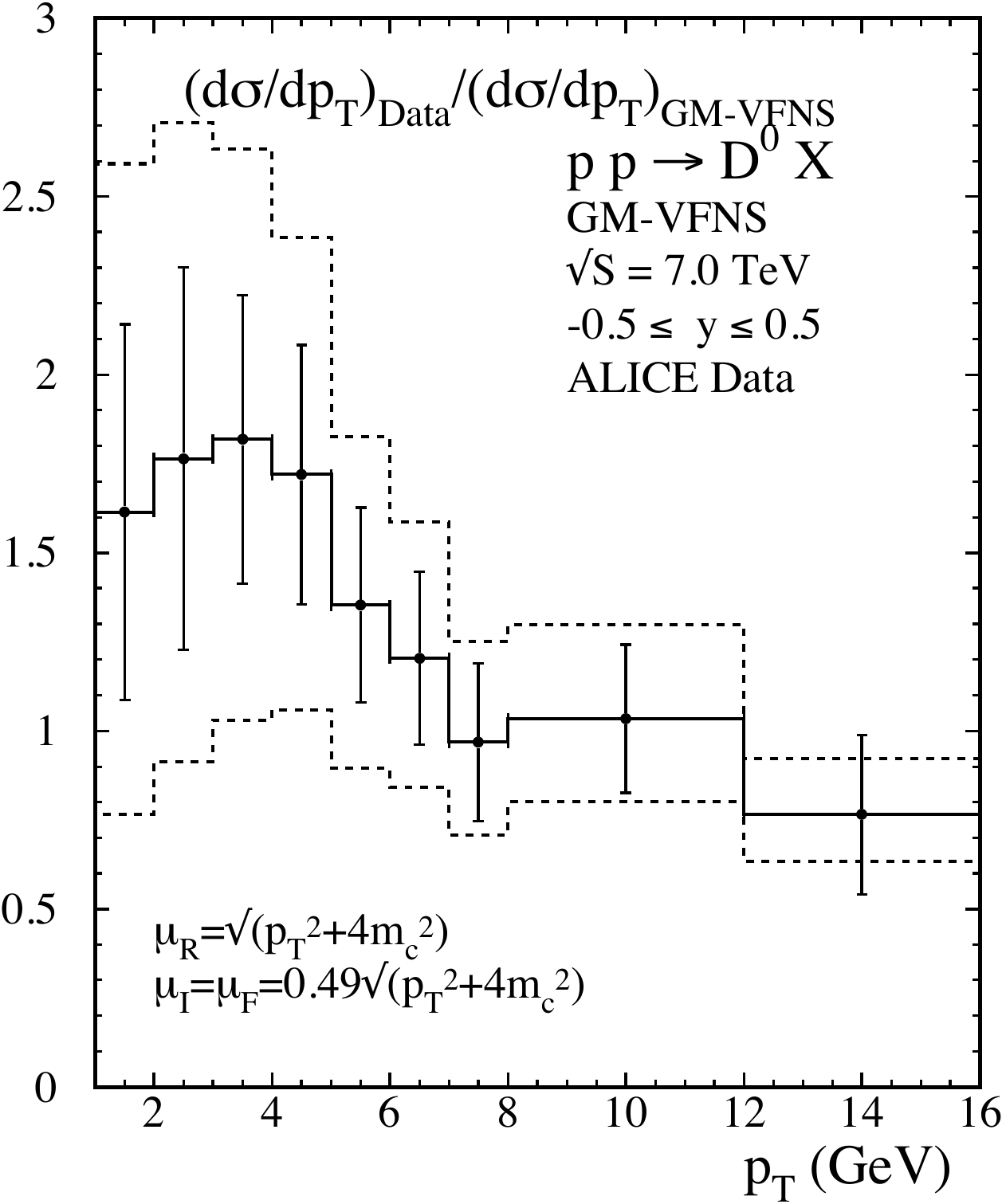}
}
\end{center}
\caption{
\label{fig:1} 
Left panel: Differential production cross section $d\sigma/dp_T$ 
of prompt $D^0$ mesons in p-p collisions at $\sqrt{S}=7$ TeV
with $|y| < 0.5$ in the $p_T$ interval $1 < p_T < 16$ GeV 
compared to ALICE data \cite{Adam:2016ich,ALICE:2011aa}. 
The data point for the bin $1 < p_T < 2$ GeV is from the 
analysis \cite{Adam:2016ich}, while the data points for 
$2 < p_T < 16$ GeV are taken from \cite{ALICE:2011aa}. The 
theoretical cross sections are calculated in the GM-VFNS with 
default scales $\mu_R = \sqrt{4m_c^2 + p_T^2}$ and $\mu_I = 
\mu_F = 0.49\sqrt{4m_c^2 + p_T^2}$. The upper and lower dashed 
histograms are calculated with $\mu_R$ changed by factors $1/2$ 
and $2$. 
Right panel: Ratios (see text) of the ALICE data over theory 
predictions. 
}
\end{figure*}
%%%%%%%%%%%%%%%%%%%%%%%%%%%%%%%%%

Before we apply this scale choice for a comparison with 
the ALICE data in p-Pb collisions at small $p_T$ 
\cite{Adam:2016ich}, we have a look at the reference p-p 
cross section. The most precise data for the 
$p_T$-differential cross section of prompt $D^0$ meson 
production at $\sqrt{S} = 7$ TeV was obtained by a combination 
of measurements without decay-vertex reconstructed in the 
low-$p_T$ range, $0 < p_T < 2$ GeV, and an analysis using 
information from decay-vertex reconstruction at larger $p_T$, 
$2 < p_T < 16$ GeV. In all cases, the rapidity is restricted 
to the range $|y| < 0.5$ and contributions from the $b \to 
D^0$ transition have been subtracted. Data and results from 
the GM-VFNS are shown in Fig.~\ref{fig:1} (left panel). 
The agreement with the default scale is very good in the 
large $p_T$ range, $p_T > 6$ GeV, and for all $p_T$ values 
the data points lie inside the theoretical range obtained 
from the scale variation of $\mu_R$. 

The ratio of data for $d\sigma/dp_T$ normalized to our 
prediction in the GM-VFNS with the {\em modified} scale 
choice is shown in the right panel of Fig.\ \ref{fig:1} 
(full-line histogram). For the larger $p_T \geq 6$ GeV 
the ratio is equal to one within the experimental accuracy. 
This is consistent with the prediction of the {\em original}€ 
scale choice, for which the same ratio was shown in Ref.\ 
\cite{Adam:2016ich} for $p_T \geq 3$ GeV. For the smaller 
$p_T$, $1 < p_T < 6$ GeV, the ratio in Fig.~\ref{fig:1} 
(right panel) increases to approximately 1.5. This is very 
similar to results based on the FONLL approach 
\cite{Cacciari:2012ny} and on the LO $k_T$ factorization 
calculation \cite{Maciula:2013wg}, which was also shown 
in \cite{Adam:2016ich}. The dashed-line histograms in the 
right panel of Fig.\ \ref{fig:1} show the ratios of the 
same data, but normalized to the GM-VFNS prediction with 
$\mu_R$ varied by factors 1/2 and 2. In order to keep the 
plot readable, we do not display the error bars for the 
experimental uncertainties in this case. The band between 
the dashed histograms thus represents the scale uncertainty 
of the ratio $d\sigma_{\rm Data}/d\sigma_{\rm GM-VFNS}$. 
We observe that inside the scale variation this ratio is 
compatible with one. 

%%%%%%%%%%%%%%%%%%%%%%%%%%%%%%%%%
\begin{figure*}[b!]
\begin{center}
\includegraphics[width=7.5cm]{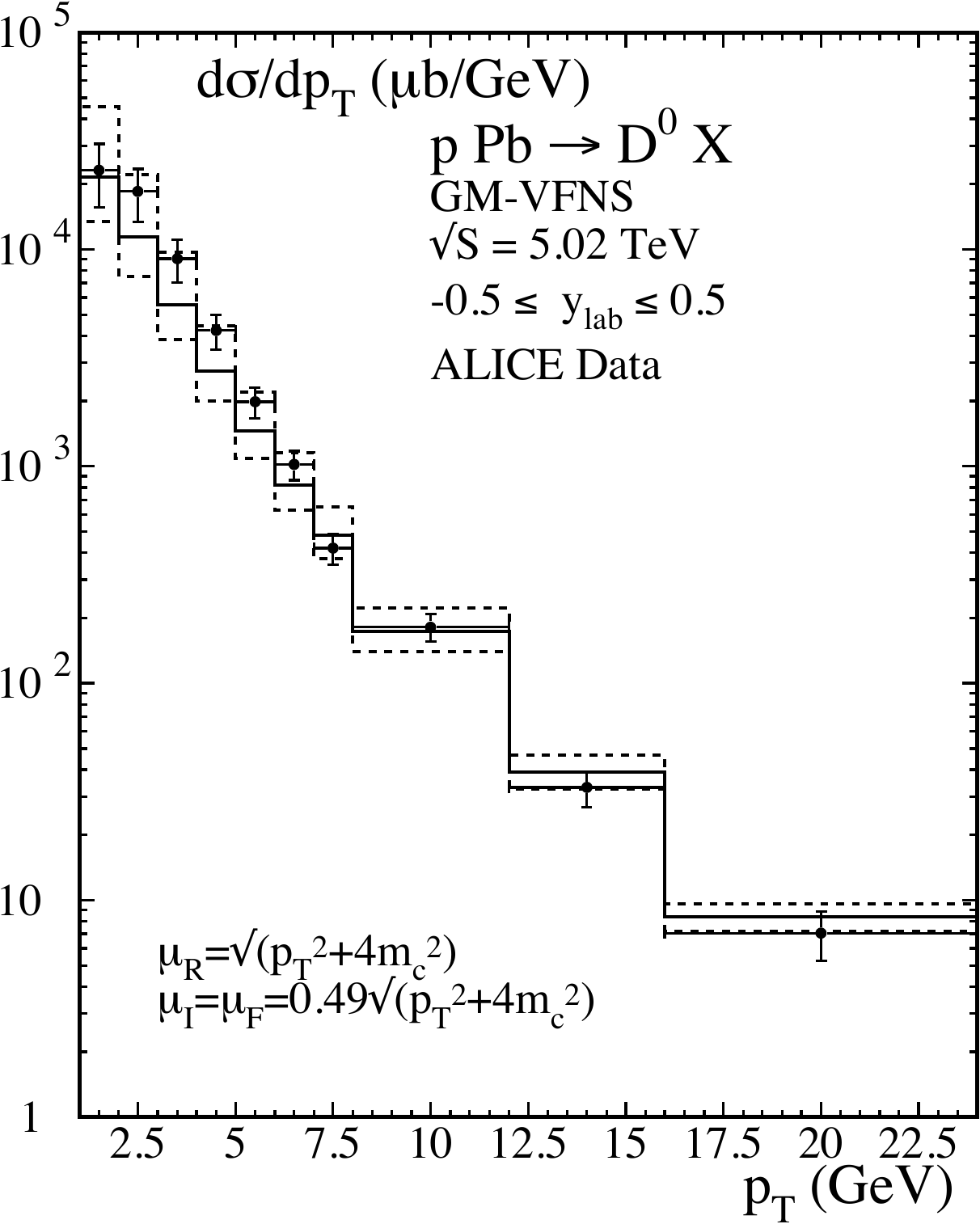}
\raisebox{-0.8mm}{
\includegraphics[width=7.6cm]{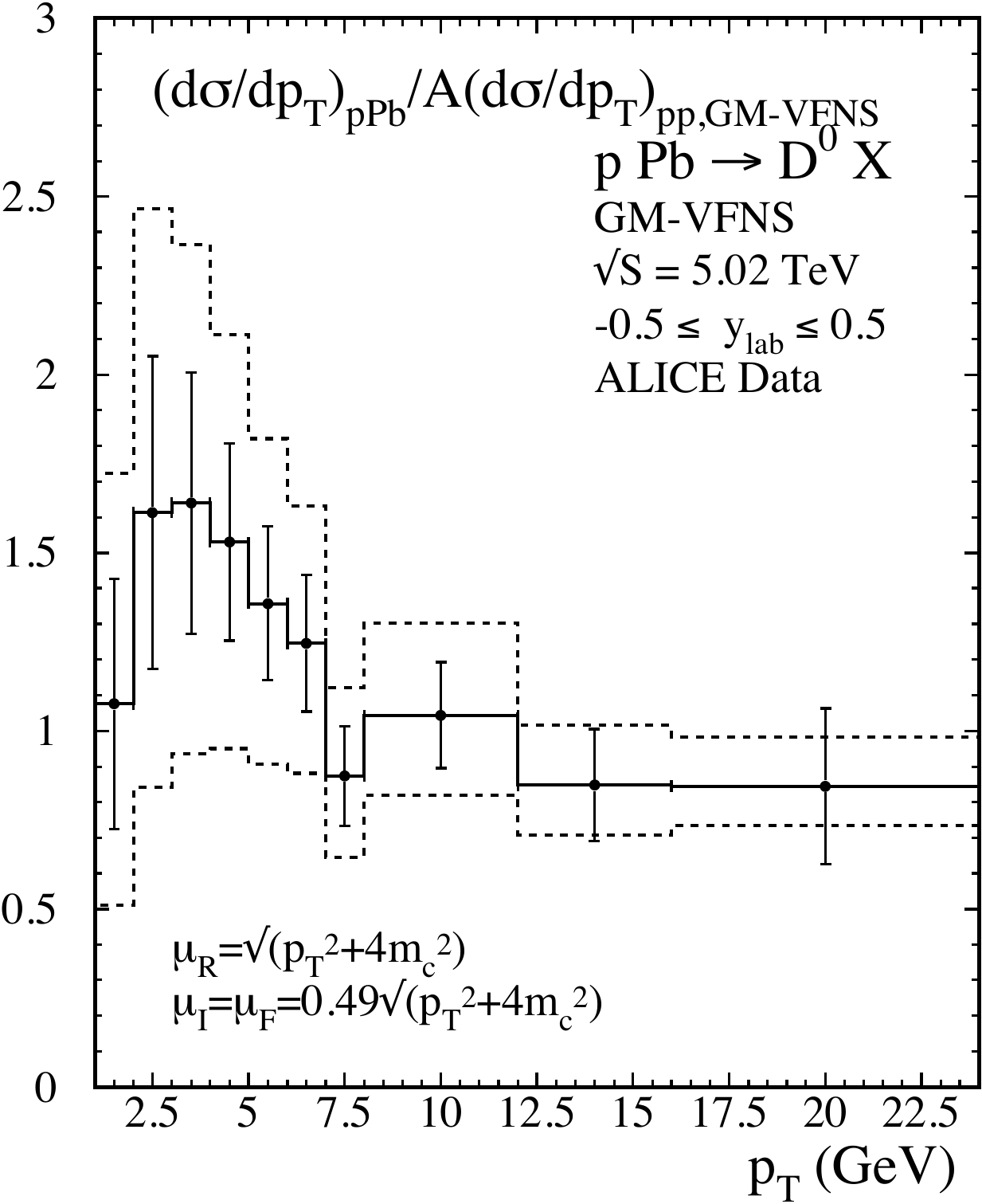}
}
\end{center}
\caption{
\label{fig:2} 
Left panel: Differential production cross section $d\sigma/dp_T$ 
of prompt $D^0$ mesons in p-Pb collisions at $\sqrt{S} = 5.02$ 
TeV with $|y| < 0.5$ of ALICE data \cite{Adam:2016ich} compared 
to $A$ times the respective p-p reference cross section calculated 
in the GM-VFNS with default scales $\mu_R = \sqrt{4m_c^2 + p_T^2}$ 
and $\mu_I = \mu_F = 0.49\sqrt{4m_c^2 + p_T^2}$. The upper and 
lower dashed histograms are calculated with $\mu_R$ changed by 
factors $1/2$ and $2$. 
Right panel: Ratios of the ALICE data over theory predictions.
}
\end{figure*}
%%%%%%%%%%%%%%%%%%%%%%%%%%%%%%%%%

%%%%%%%%%%%%%%%%%%%%%%%%%%%%%%%%%
\begin{figure*}
\begin{center}
\includegraphics[width=7.5cm]{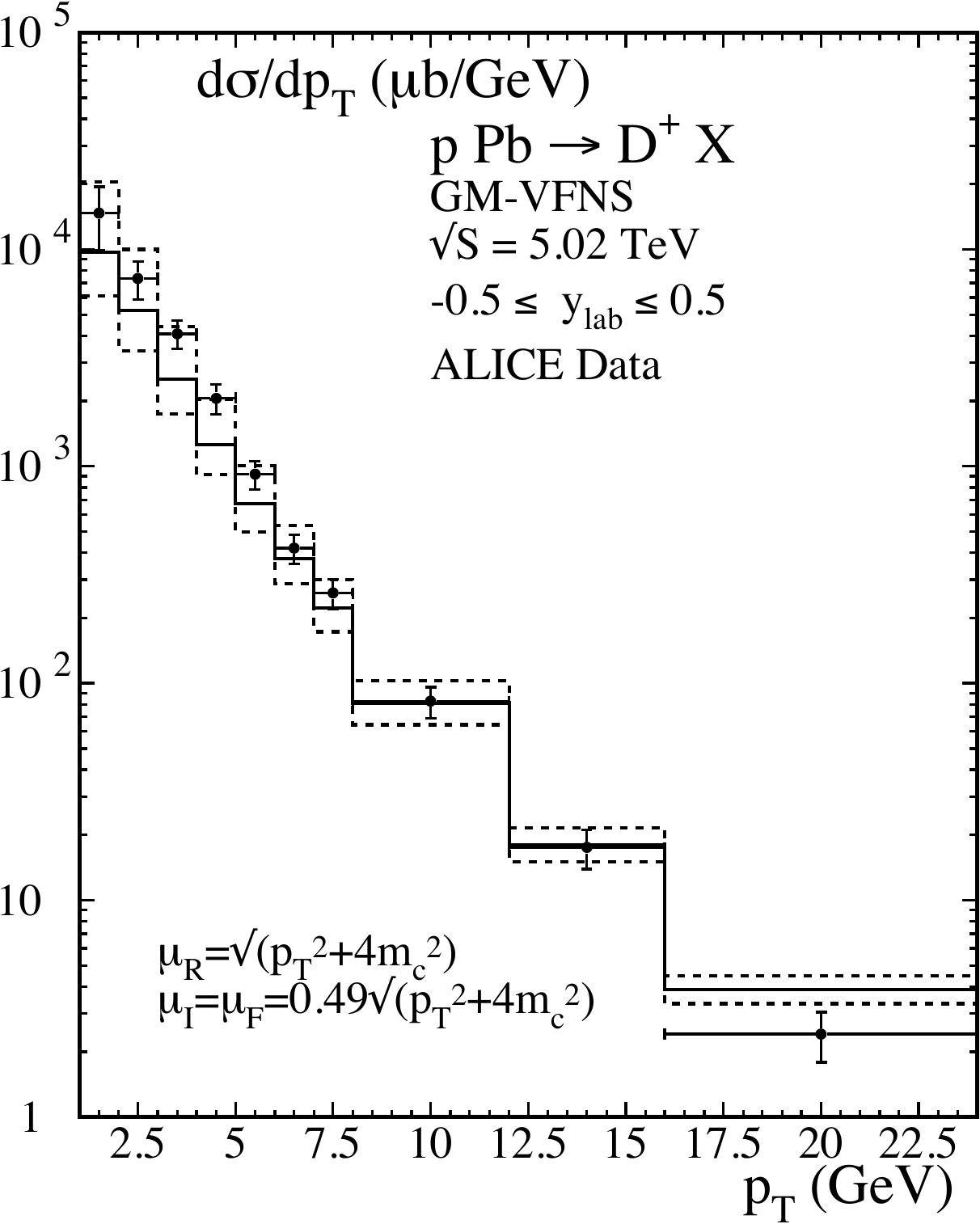}
\raisebox{-0.8mm}{
\includegraphics[width=7.6cm]{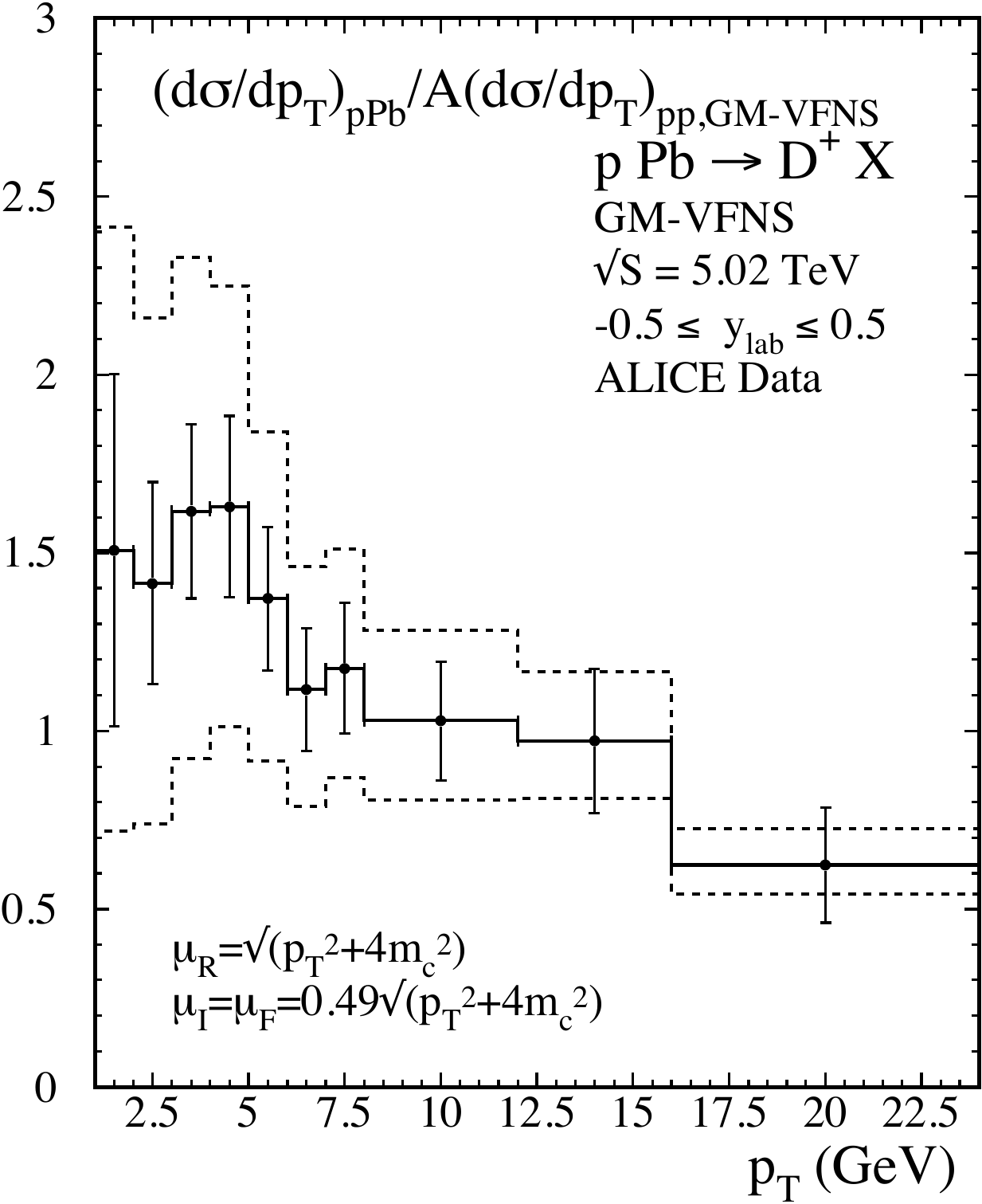}
}
\end{center}
\caption{
\label{fig:3} 
Left panel: Differential production cross section $d\sigma/dp_T$ 
of prompt $D^+$ mesons in p-Pb collisions at $\sqrt{S} = 5.02$ 
TeV with $|y| < 0.5$ of ALICE data \cite{Adam:2016ich} 
compared to $A$ times the respective p-p reference cross section 
calculated in the GM-VFNS with default scales $\mu_R = 
\sqrt{4m_c^2 + p_T^2}$ and $\mu_I = \mu_F = 0.49\sqrt{4m_c^2 + p_T^2}$. 
The upper and lower dashed histograms are calculated with $\mu_R$ 
changed by factors $1/2$ and $2$. 
Right panel: Ratios of the ALICE data over theory predictions. 
} 
\end{figure*}
%%%%%%%%%%%%%%%%%%%%%%%%%%%%%%%%%

%%%%%%%%%%%%%%%%%%%%%%%%%%%%%%%%%
\begin{figure*}
\begin{center}
\includegraphics[width=7.5cm]{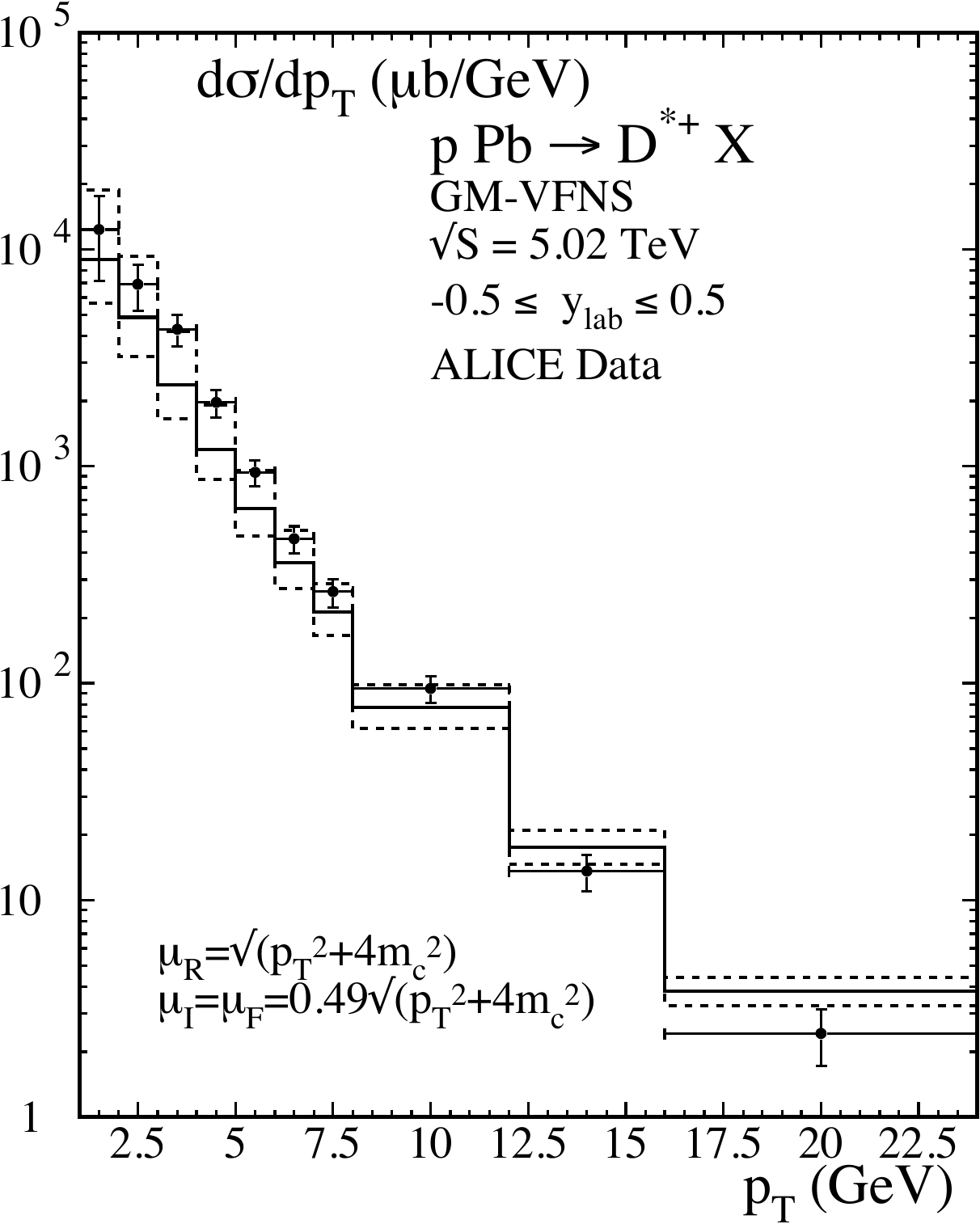}
\raisebox{-0.8mm}{
\includegraphics[width=7.6cm]{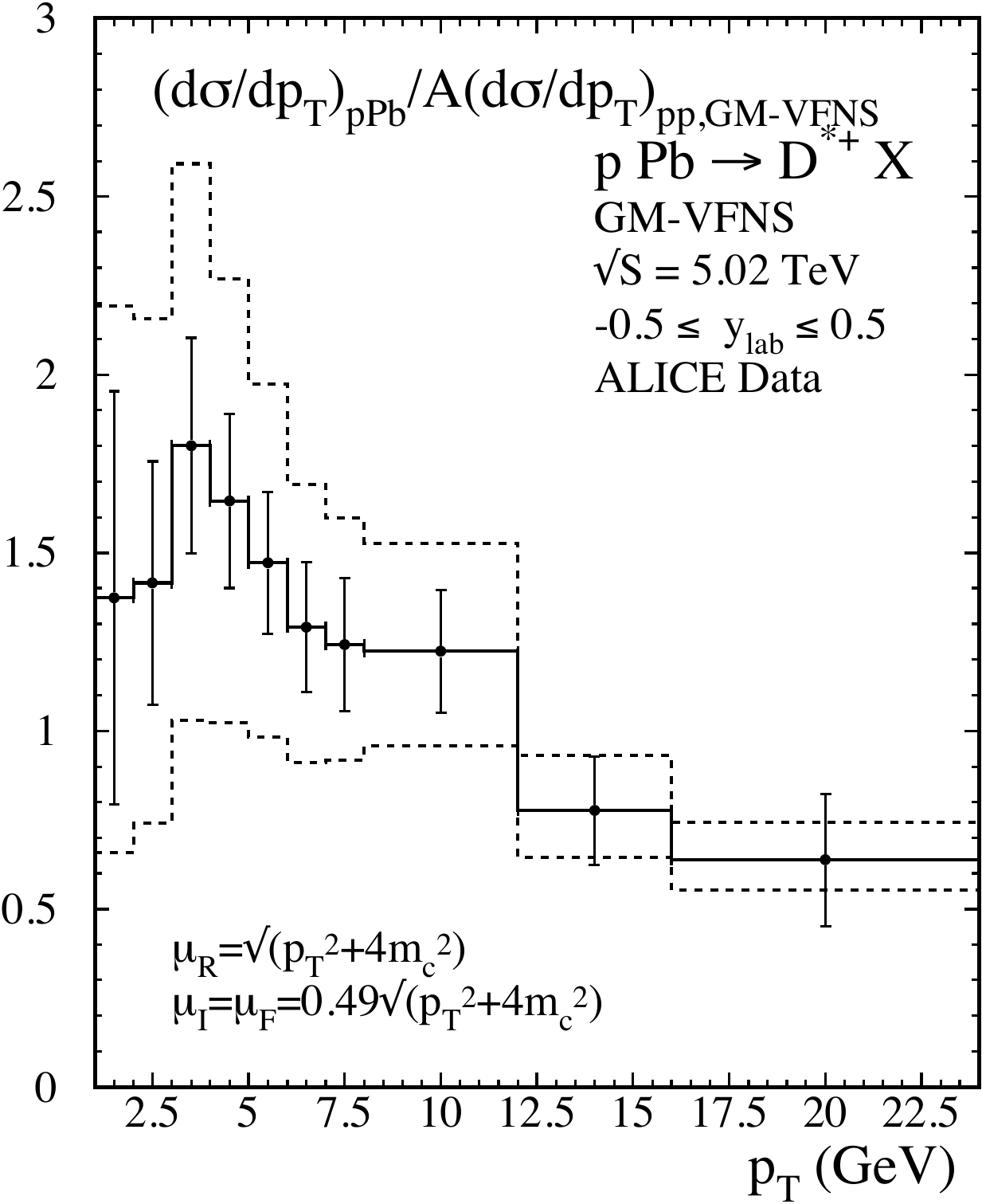}
}
\end{center}
\caption{
\label{fig:4} 
Left panel: Differential production cross section $d\sigma/dp_T$ of 
prompt $D^{*+}$ mesons in p-Pb collisions at $\sqrt{S} = 5.02$ TeV 
with $|y| < 0.5$ of ALICE data \cite{Adam:2016ich} compared to 
$A$ times the respective p-p reference cross section calculated in 
the GM-VFNS with default scales $\mu_R =\sqrt{4m_c^2 + p_T^2}$ and 
$\mu_I = \mu_F = 0.49\sqrt{4m_c^2 + p_T^2}$. The upper and lower 
dashed histograms are calculated with $\mu_R$ changed by factors 
$1/2$ and $2$. 
Right panel: Ratios of the ALICE data over theory predictions. 
} 
\end{figure*}
%%%%%%%%%%%%%%%%%%%%%%%%%%%%%%%%%

%%%%%%%%%%%%%%%%%%%%%%%%%%%%%%%%%
\begin{figure*}
\begin{center}
\includegraphics[width=7cm]{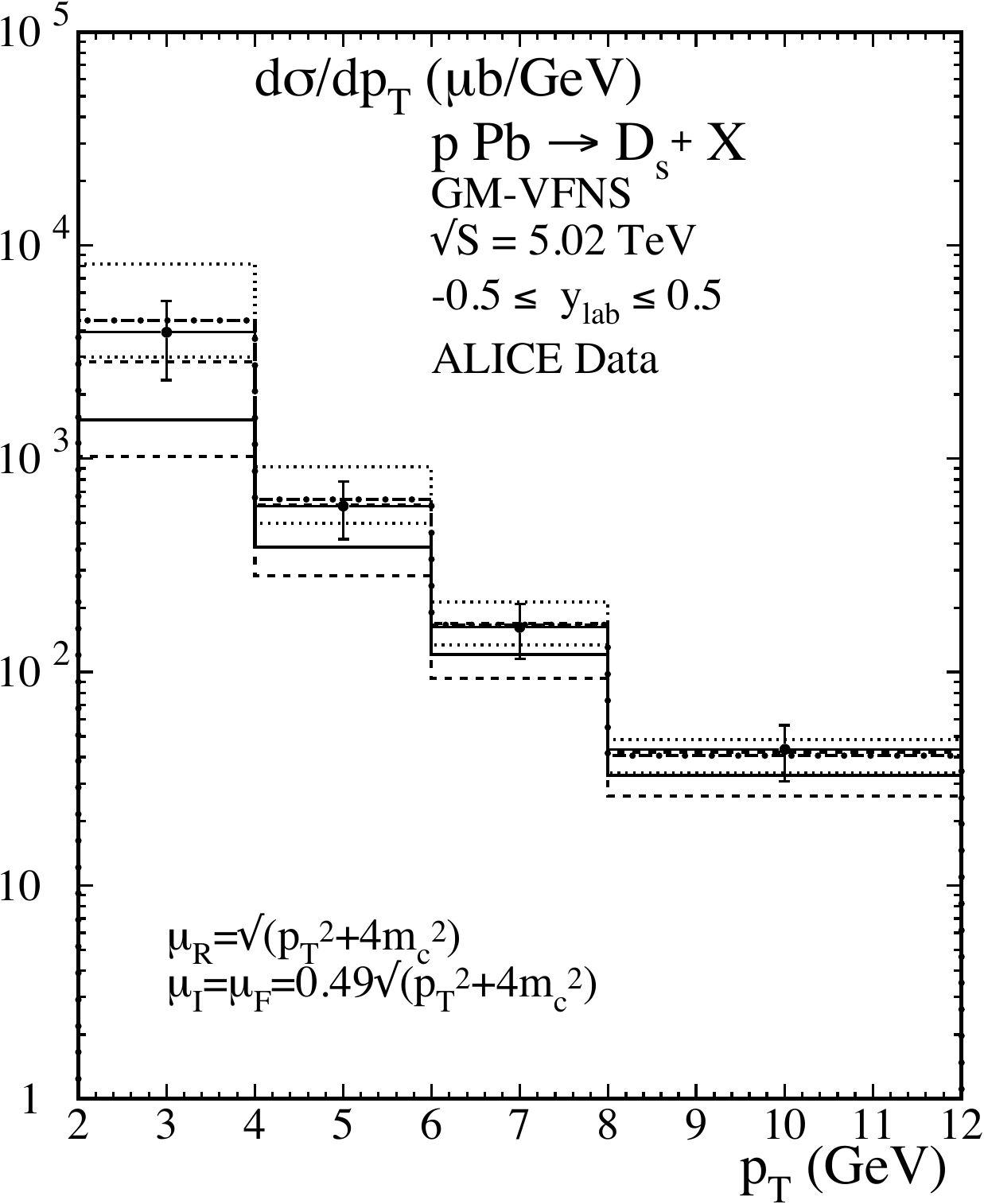}
\\
\includegraphics[width=7cm]{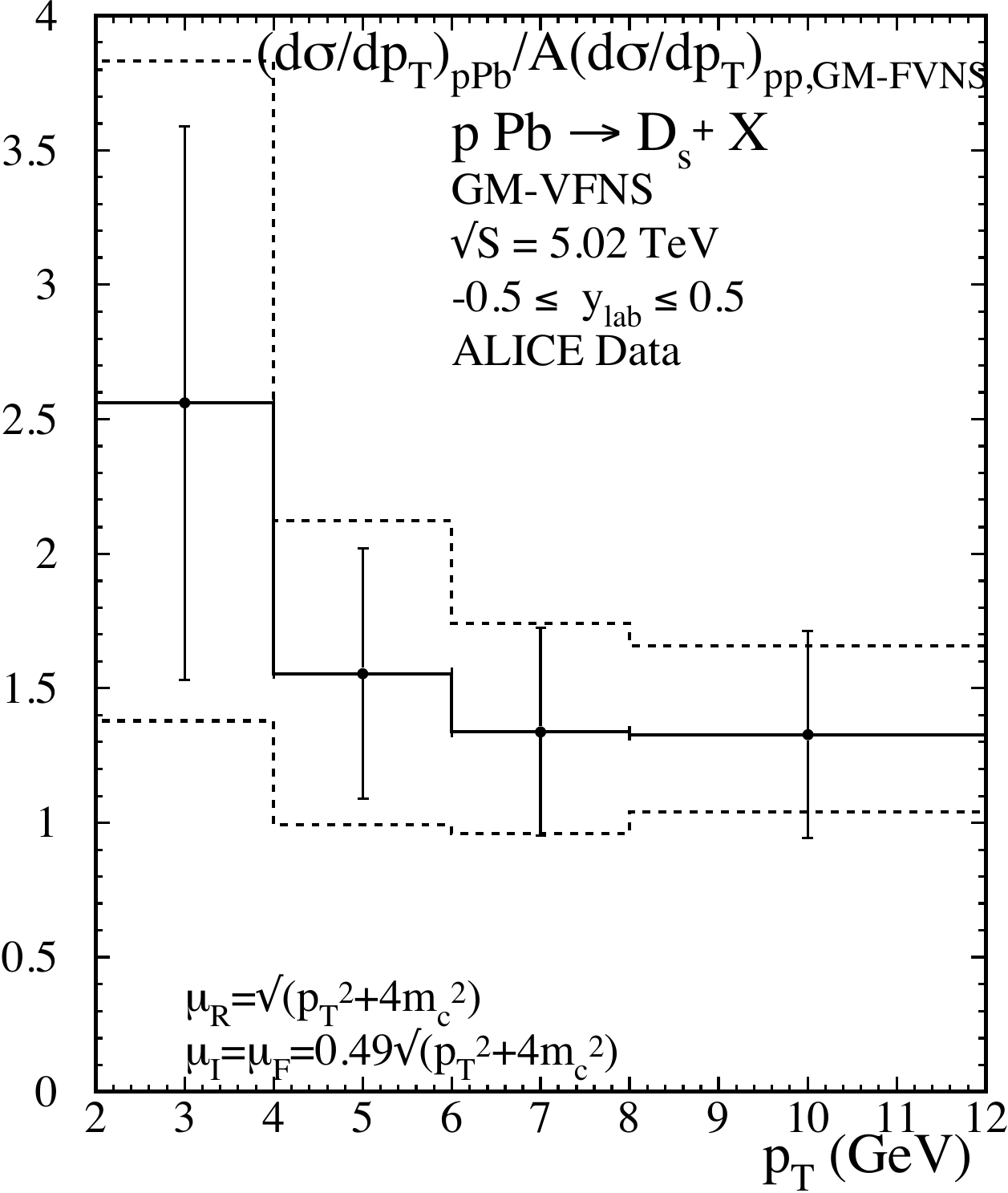}
\raisebox{-2mm}{
\includegraphics[width=7cm]{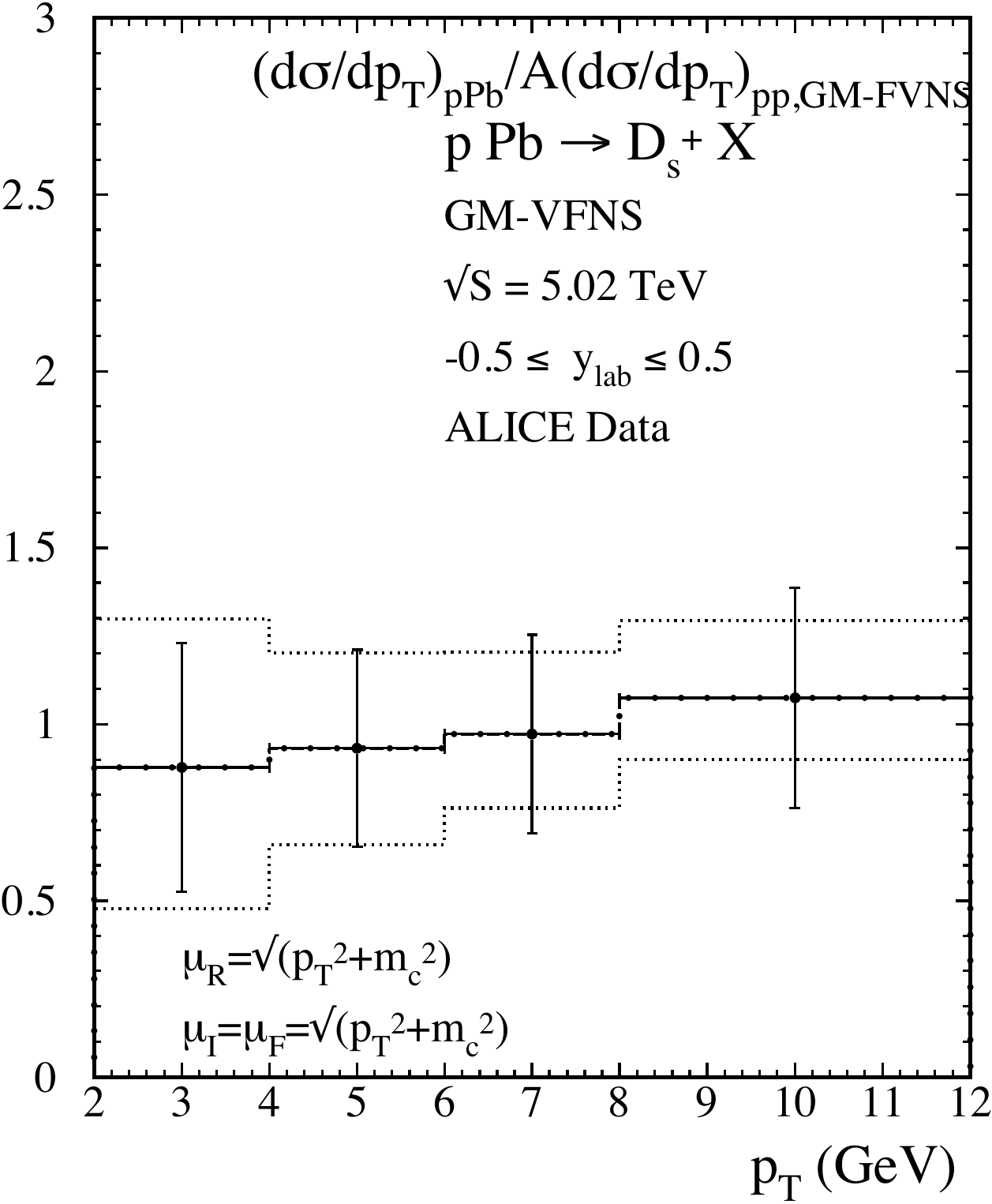}
}
\end{center}
\caption{
\label{fig:5} 
Upper panel: Differential production cross section $d\sigma/dp_T$ 
of prompt $D_s^+$ mesons in p-Pb collisions at $\sqrt{S} = 5.02$ 
TeV with $|y| < 0.5$. We compare ALICE data \cite{Adam:2016ich} 
with $A$ times the respective p-p reference cross section 
calculated in the GM-VFNS with default scales $\mu_R = 
\sqrt{4m_c^2 + p_T^2}$ and $\mu_I = \mu_F = 0.49\sqrt{4m_c^2 + p_T^2}$. 
The upper and lower dashed histograms are calculated with $\mu_R$ 
changed by factors $1/2$ and $2$. The dashed-dotted histogram is 
obtained for the {\em original}€ scale choice and the light dotted 
histograms for its corresponding scale variations.
Lower panels: Ratios of the ALICE data over theory predictions 
for the {\em modified} scale choice (left) and the {\em original} 
scale choice (right).
} 
\end{figure*} 
%%%%%%%%%%%%%%%%%%%%%%%%%%%%%%%%%

Now we continue with a comparison of theory predictions and 
ALICE data for p-Pb collisions. Theoretical predictions are 
obtained from the p-p cross section by multiplication with 
the mass number $A = 208$, $Ad\sigma/dp_T$. Data are available 
at $\sqrt{S} = 5.02$ TeV in the rapidity region $|y| < 0.5$. 
Our results in the GM-VFNS with the {\em modified} scale 
choice are shown in Figs.\ \ref{fig:2}, \ref{fig:3}, \ref{fig:4}, 
and \ref{fig:5} (left panels) for $D^0$, $D^+$, $D^{*+}$ and $D_s^+$ 
production, in each case together with the data from 
\cite{Adam:2016ich} as a function of $p_T$ for bins in the range 
$1 < p_T < 24$ GeV. Except for two points at the largest $p_T$ 
(see Figs.\ \ref{fig:3} and \ref{fig:4}) the error bars of the 
data points overlap with the uncertainty range due to scale 
variations. As for p-p collisions, the ALICE data shown in Fig.\ 
\ref{fig:2} are obtained for prompt $D^0$ production in the interval 
$0 < p_T < 2$ GeV (only data for $p_T > 1$ GeV are shown) without 
decay-vertex reconstruction \cite{Adam:2016ich} and for $p_T > 2$ 
GeV with decay-vertex reconstruction \cite{Abelev:2014hha}. The 
data for the other three $D$-meson species $D^+$, $D^{*+}$ and 
$D_s^+$ are taken from Ref.\ \cite{Abelev:2014hha}.

%%%%%%%%%%%%%%%%%%%%%%%%%%%%%%%%%
\begin{figure*}[b!]
\begin{center}
\includegraphics[width=7cm]{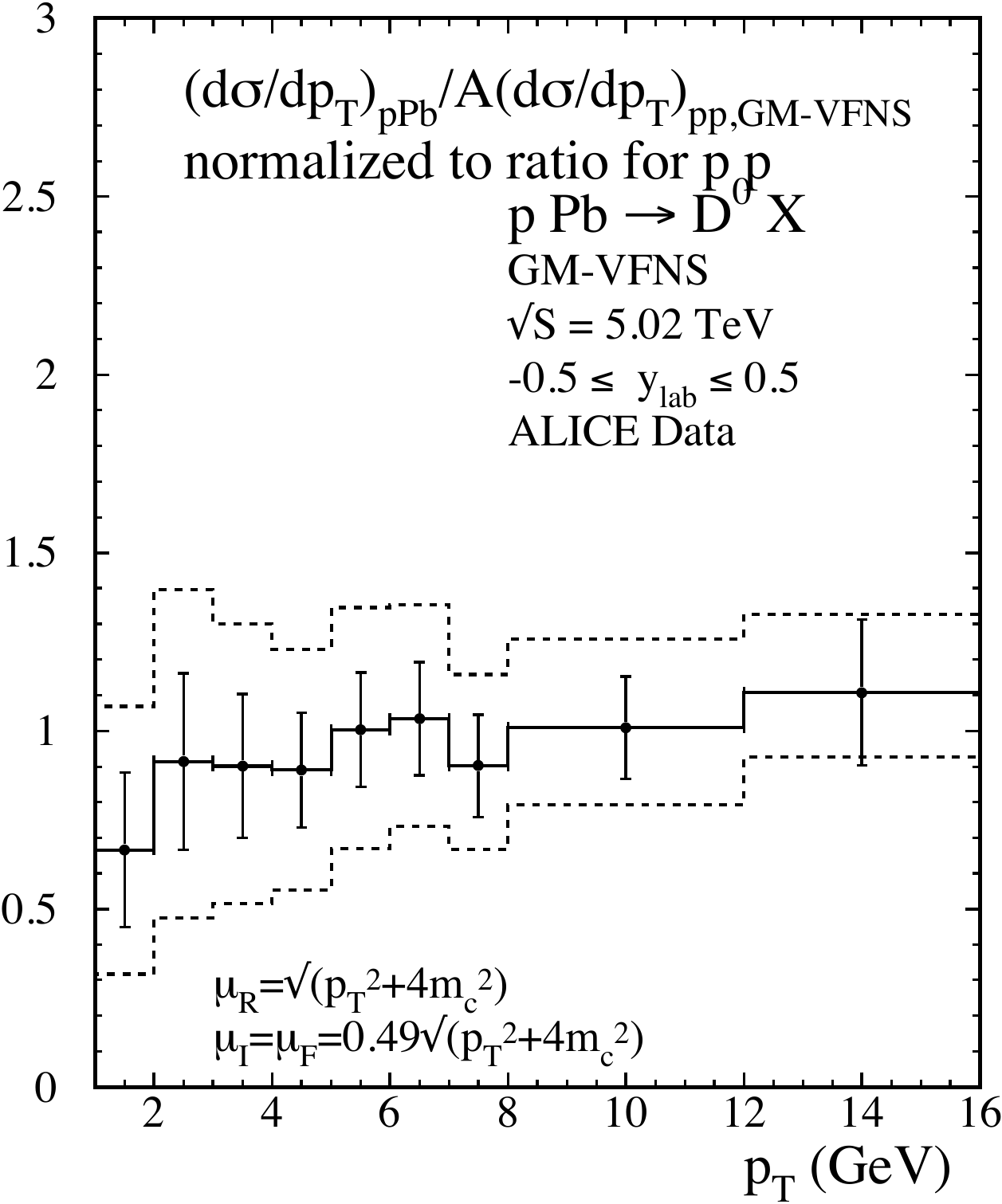}
\end{center}
\caption{
The ratio of ALICE data for $D^0$ production in p-Pb collisions 
at $\sqrt{S} = 5.02$ TeV over theory, normalized to the $D^0$ 
data in p-p collisions (see text). The error bars show the 
uncertainty of the p-Pb data and the band of dashed histograms 
represents the theory uncertainty due to variations of the 
renormalization scale.
\label{fig:6} 
} 
\end{figure*} 
%%%%%%%%%%%%%%%%%%%%%%%%%%%%%%%%%

Corresponding ratios for ALICE data normalized to our theoretical 
results for $Ad\sigma/dp_T$ are presented in the right panels of 
Figs.\ \ref{fig:2}, \ref{fig:3}, \ref{fig:4}, and the lower 
panels of \ref{fig:5}. Again, we decide to present scale 
uncertainties by normalizing the data to varied theory 
predictions with $\mu_R$ scaled up and down by factors of 
1/2 and 2 (dashed-line histograms) and show the ratios 
$R_\pm = d\sigma_{\rm p-Pb, data} / 
(A d\sigma_{\rm p-p, GM-VFNS}(\mu_\pm))$ where $\mu_\pm$ 
denotes the varied renormalization scale. The band enclosed 
by $R_\pm$ should contain unity if there is a scale choice 
which leads to agreement between theory and experiment. This 
is indeed the case, except for $D^+$ and $D^{\ast +}$ 
production at the largest $p_T$ where the ratio falls 
slightly below one. 

The shape of the $p_T$ dependence of the ratios looks rather 
similar for all cases, compare for example the case of $D^0$ 
production for p-Pb collisions at $\sqrt{S} = 5.02$ TeV in 
Fig.~\ref{fig:2} and for p-p collisions at $\sqrt{S} = 7$ TeV 
in Fig.~\ref{fig:1}. The similarity between p-p and p-Pb 
collisions is even more clearly visible when we consider the 
ratios of the results shown in the right panels of Figs.\ 
\ref{fig:1} and \ref{fig:2}. This is done in Fig.\ \ref{fig:6} 
where we show 
\begin{eqnarray*}
R_i & = &
\left[
\frac{d\sigma_{\rm pPb,data}(\sqrt{s}=5)}%
{A d\sigma_{\rm pp,data}(\sqrt{s}=7))}
\cdot
\frac{d\sigma_{\rm pp, GM-VFNS}(\mu_0, \sqrt{s}=7)}%
{d\sigma_{\rm pp, GM-VFNS}(\mu_0, \sqrt{s}=5)}
\right]
\times  
\frac{d\sigma_{\rm pp, GM-VFNS}(\mu_0, \sqrt{s}=5)}%
{d\sigma_{\rm pp, GM-VFNS}(\mu_i, \sqrt{s}=5)}
\end{eqnarray*}
where $\mu_{i}$ denotes the renormalization scale varied 
up and down by factos of 1/2 and 2 around its central value 
$\mu_0$. The first factor in brackets is represented by 
the full histogram in Fig.\ \ref{fig:6}. It is the ratio 
of p-Pb over p-p data, properly normalized to the same 
value of $\sqrt{S}$ using the GM-VFNS prediction. The 
error bars shown here represent the uncertainty of the p-Pb 
data only. The band of dashed-line histograms represents 
an estimate of the scale uncertainty, evaluated at 
$\sqrt{s} = 5.02$ TeV (see the last factor in the definition 
of $R_i$ given above). Since $R_i \equiv 1$ is contained 
inside this band we conclude that the data do not require 
corrections, for example due to initial-state interactions 
in the Pb nucleus. 

For the other mesons, $D^+$, $D^{*+}$ and $D_s^+$ in Figs.\ 
\ref{fig:3}, \ref{fig:4}, and \ref{fig:5} the pattern of ratios 
looks also quite similar. For the larger $p_T$ bins the ratio 
is equal to one within errors, and for the smaller $p_T$ bins 
the ratio is close to 1.5. We remark that the nuclear 
modification factor $R_{\rm pPb}$ is consistent with one for 
all four $D$ meson species if the theoretical uncertainty due 
to scale variations is taken into account.

We can compare our results with the nuclear modification factor 
presented in Ref.\ \cite{Adam:2016ich}. The ratios $R_{\rm pPb}$ 
for $D^0$, $D^+$ and $D^{*+}$ given there are much closer to 
one than our calculated ratios shown in Figs.\ \ref{fig:2}, 
\ref{fig:3}, and \ref{fig:4}. Note that the p-p cross sections 
used in Ref.\ \cite{Adam:2016ich} to obtain the ratios 
$R_{\rm pPb}$ have been deduced from the measured cross 
sections at $\sqrt{S} = 7$ TeV by extrapolation to $\sqrt{S} 
= 5.02$ TeV. It would be premature to interpret the observed 
small deviations of the nuclear modification factors from 
one as a sign of initial-state interaction effects as long 
as we see similar deviations for p-p collisions as shown 
in Fig.~\ref{fig:1}, right panel. It has been shown in Ref.\ 
\cite{Adam:2016ich} that theoretical expectations for 
deviations of $R_{\rm pPb}$ from one for several models 
existing in the literature are rather small at large $p_T$. 
Only towards small values of $p_T$ model predictions start 
to deviate from one by more than 10 percent or so. Future 
higher-precision data may allow to exclude some of the 
theoretical approaches, but right now experimental 
uncertainties are still too large to draw any firm 
conclusion. 

%%%%%%%%%%%%%%%%%%%%%%%%%%%%%%%%%
\begin{figure*}[t!]
\begin{center}
\includegraphics[width=7.1cm]{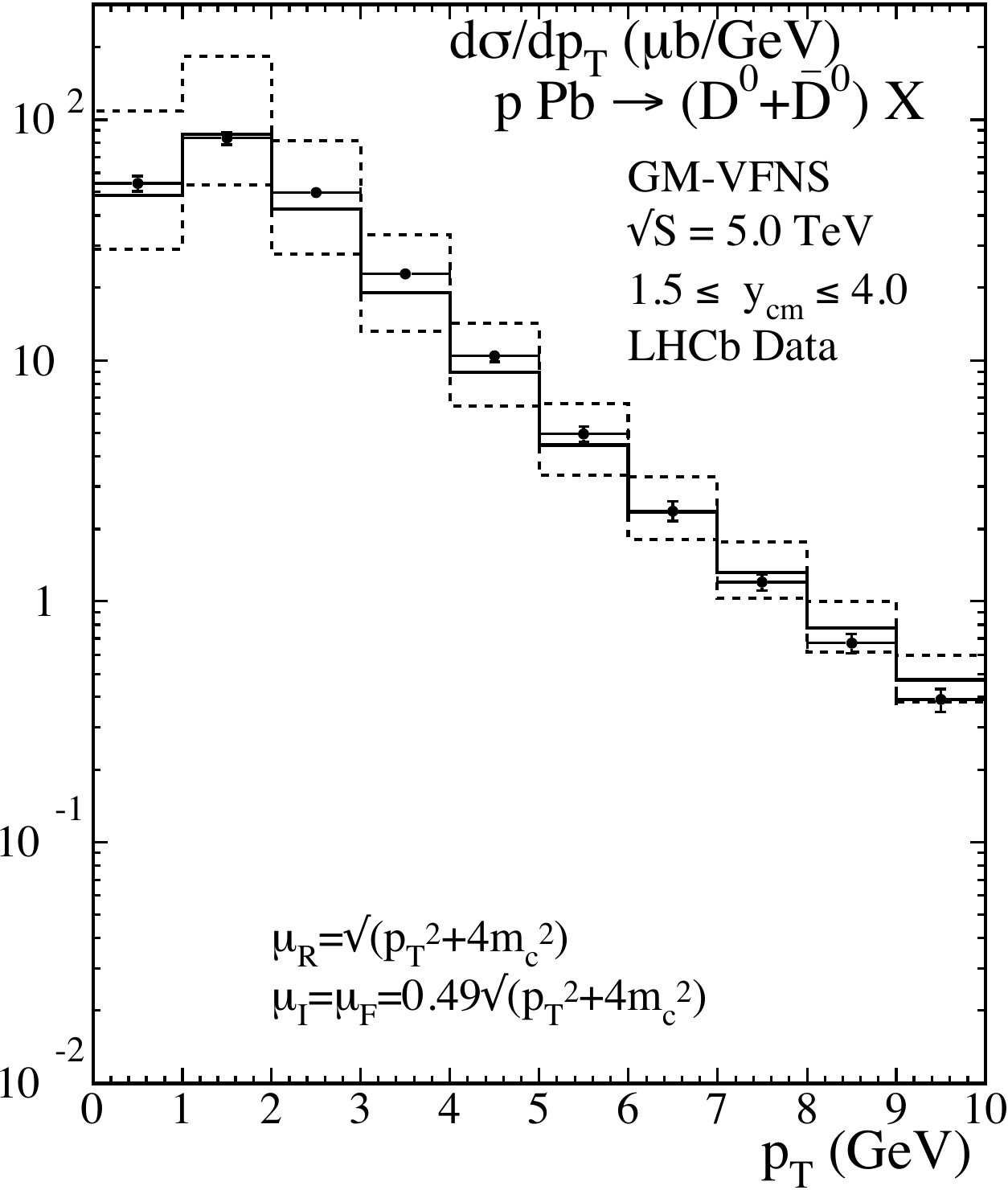}
\raisebox{-0.7mm}{
\includegraphics[width=7.1cm]{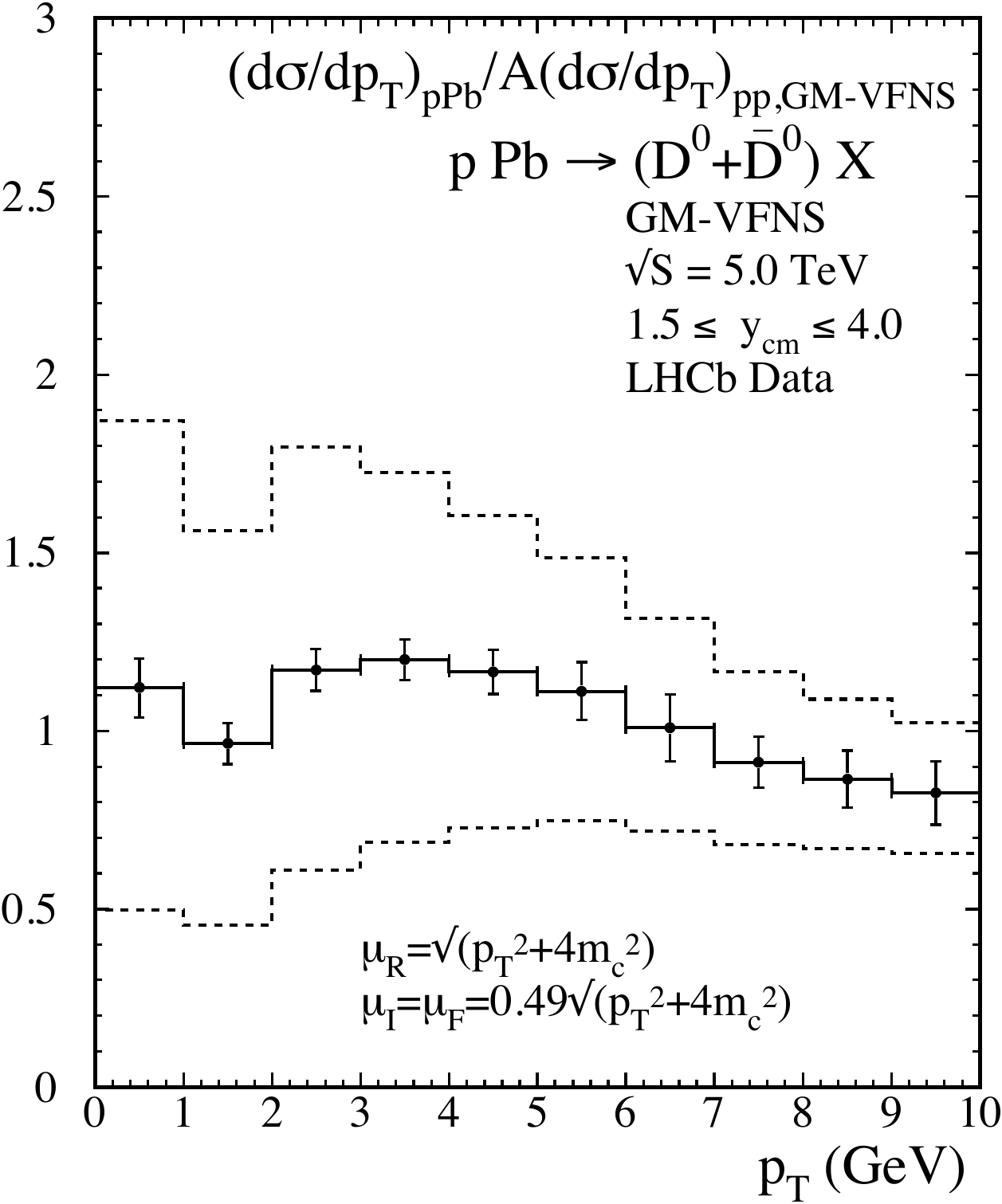}
}
\\
\includegraphics[width=7.1cm]{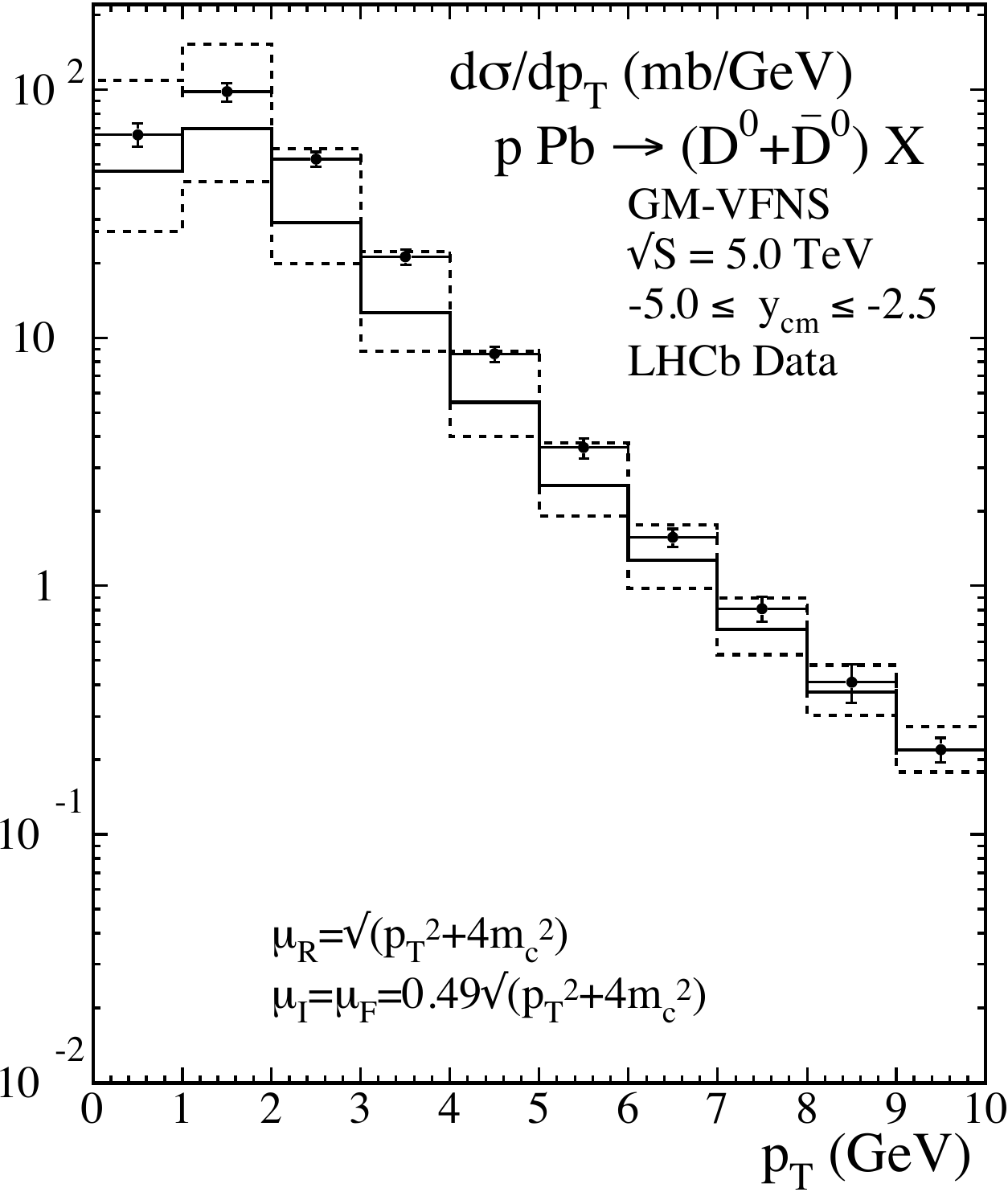}
\raisebox{-0.7mm}{
\includegraphics[width=7.1cm]{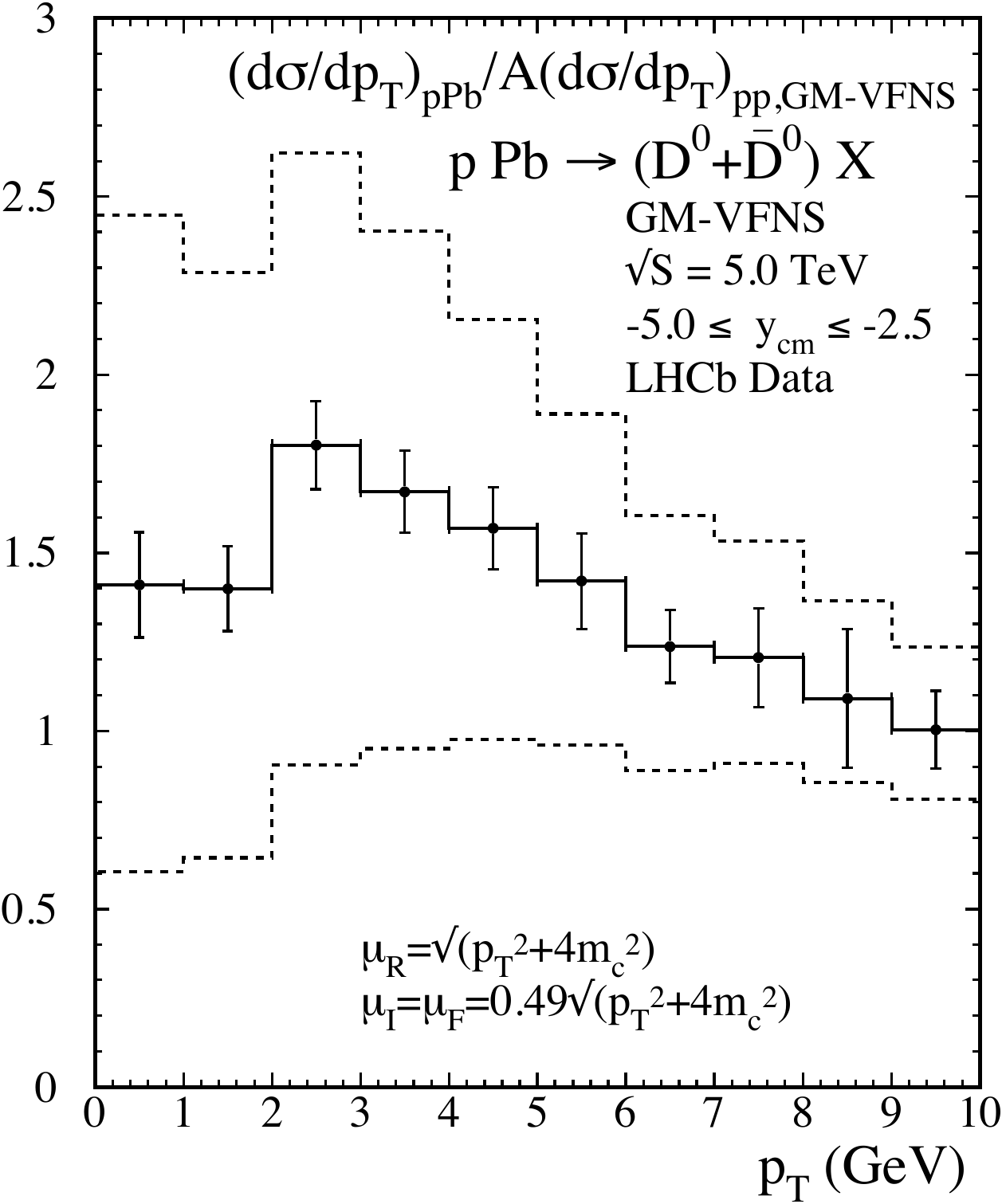}
}
\end{center}
\caption{
The $p_T$ distribution for $D^0 + \bar{D}^0$ production in 
p-Pb collisions compared with data from the LHCb collaboration 
taken at $\sqrt{S} = 5$ TeV. The left plots show the 
differential cross sections $d\sigma/dp_T$, the right 
plots show the ratios of data over theory. Full and dashed 
lines are defined as in the previous figures (see also text). 
Data and ratios in the upper part are for the forward region 
$1.5 \leq y_{\rm cm} \leq 4.0$ and in the lower part for the 
backward region $-5.0 \leq y_{\rm cm} \leq -2.5$. 
\label{fig:L6} 
} 
\end{figure*} 
%%%%%%%%%%%%%%%%%%%%%%%%%%%%%%%%%
 
Finally we compare predictions from the GM-VFNS approach with 
most recent data from the LHCb collaboration \cite{Aaij:2017gcy}. 
For p-p collisions, a rather good agreement between theory 
predictions and LHCb data for the differential cross section 
$d\sigma/dp_T$ in various rapidity bins in the forward direction 
was already observed in Ref.\ \cite{Benzke:2017yjn}. The recent 
measurements of p-Pb cross sections at LHCb \cite{Aaij:2017gcy} 
have provided us with more information about the dependence on 
the rapidity $y_{\rm cm}$ in the nucleon-nucleon centre-of-mass 
system and allow us to study the forward and backward regions 
separately. We note that experimental uncertainties are much 
smaller than for the other measurements described before. In 
Fig.\ \ref{fig:L6} we show two sets of plots, one for the 
forward region, $1.5 \leq y_{\rm cm} \leq 4.0$ (upper plots) 
and one for the backward region $-5.0 \leq y_{\rm cm} \leq 
-2.5$ (lower plots). All data points agree with theory within 
the scale uncertainty band. In the right plots of Fig.\ 
\ref{fig:L6} we show ratios of data for p-Pb collisions 
normalized to $A$ times theory predictions for p-p scattering. 
The deviation of these ratios from one are not very large in 
the forward region, but significantly above one for backward 
rapidities. We expect that this observation can be explained 
by using appropriately chosen nuclear PDFs. At present, 
nuclear PDFs have very large errors \cite{deFlorian:2011fp, 
Kovarik:2015cma,Eskola:2017rmp} and a direct comparison with 
the available nPDFs is not very instructive. However, one can 
conclude that these precise LHCb data will help to narrow down 
possible nPDF parametrizations. We note that the forward-backward 
ratio discussed in the LHCb publication will be particularly 
interesting for a study of nuclear PFFs since it is not 
affected by large scale uncertainties.

%%%%%%%%%%%%%%%%%%%%%%%%%%%%%%%%%%%%%%%%%%%%%%%%%%%%%%%%%%%%%%%%%%%%%%%
\section{{\boldmath $B$} meson production in p-Pb collisions}

Up to now, cross section data of $d\sigma/dp_T$ for $B$-meson 
production ($B^+$, $B^0$ and $B_s^0$) in p-Pb collisions at 
$\sqrt{S} = 5.02$ TeV are available only for larger $p_T$ 
values above 10 GeV \cite{Khachatryan:2015uja}, in the range 
$10 < p_T < 60$ GeV. In Ref.\ \cite{Khachatryan:2015uja} data 
have been compared with $A$ times the FONLL prediction for p-p 
collisions \cite{Cacciari:2012ny}. At $\sqrt{S} = 7$ TeV the 
LHCb collaboration has measured the p-p cross section $d\sigma/dp_T$ 
down to $p_T=0$ for $B^+ + B^-$, $B^0 + \bar{B}^0$ and 
$B_s^0 + \bar{B}_s^0$ production in the forward region $2 \leq y 
\leq 4.5$ \cite{Aaij:2012jd,Aaij:2013noa}. These data have been 
compared with our GM-VFNS predictions using the {\em modified} scale 
$0.5 \sqrt{m_b^2 + p_T^2}$. The comparison between the LHCb data
and our predictions showed reasonably good agreement for all 
three $B$ meson species \cite{Kniehl:2015fla}. In this reference 
we compared the GM-VFNS predictions also for $B^+$-meson 
production measured by the ATLAS collaboration 
\cite{ATLAS:2013cia} where data extend into the very large 
$p_T$-range, $9 < p_T < 120$ GeV, for various rapidity intervals 
in the range $0 < |y| < 2.25$. In this comparison we found 
agreement between data and theory except for the lowest $p_T$ 
bin, 9-13 GeV, where the data are slightly overestimated. 

%%%%%%%%%%%%%%%%%%%%%%%%%%%%%%%%%
\begin{figure*}[t!]
\begin{center}
\includegraphics[width=7.5cm]{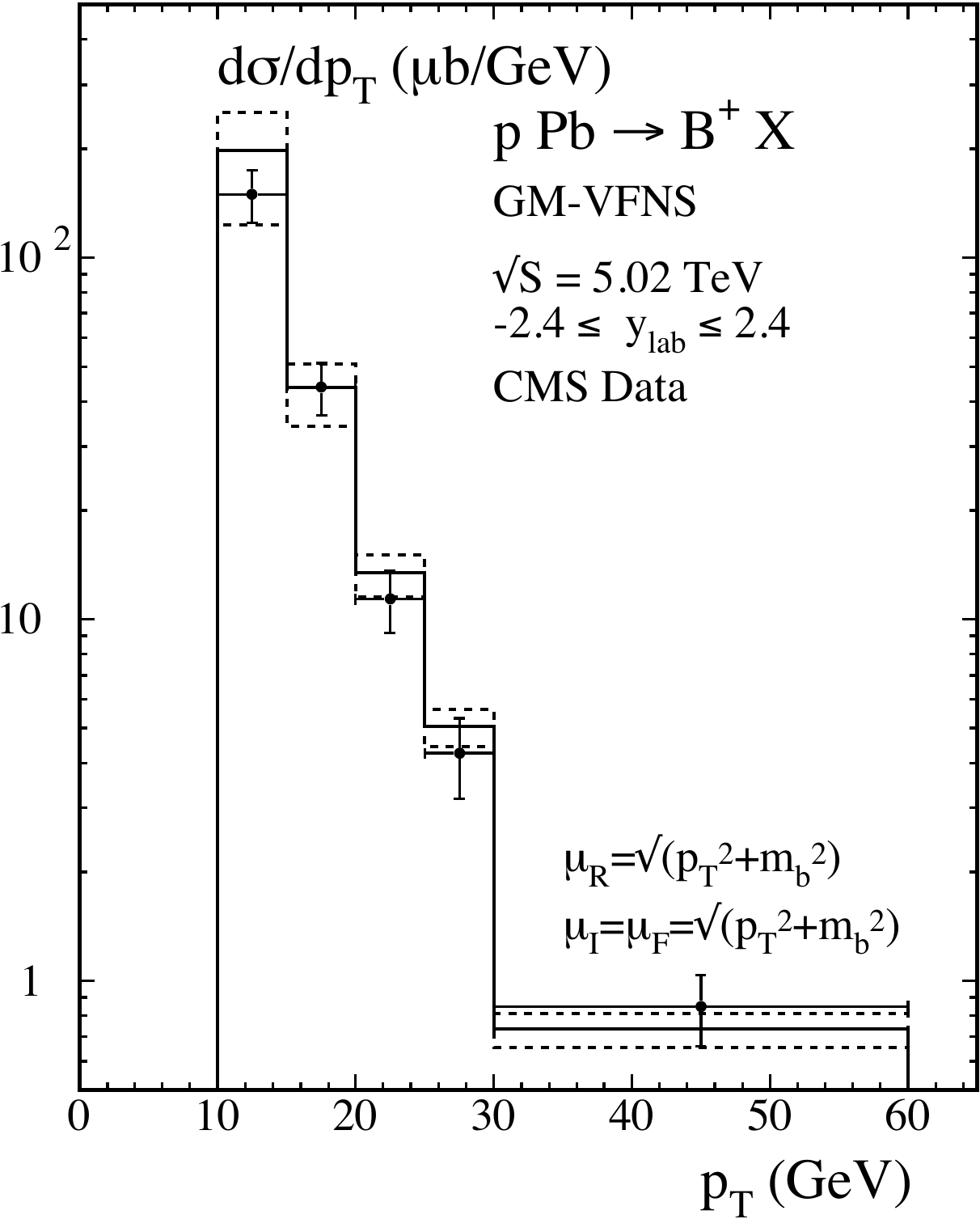}
\raisebox{1mm}{
\includegraphics[width=7.5cm]{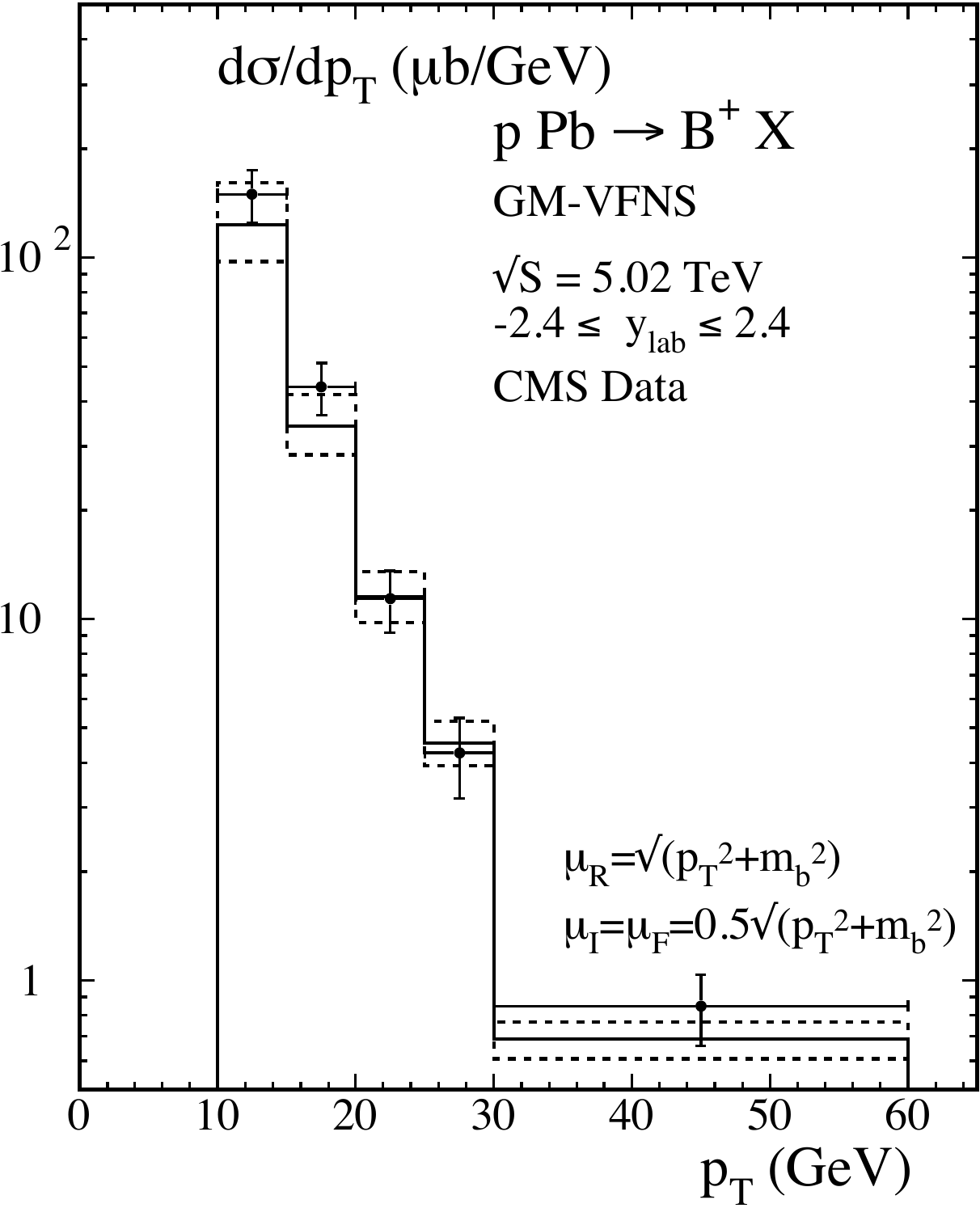}
}
\end{center}
\caption{ 
\label{fig:7} 
Differential cross section $A d\sigma/dp_T$ as a function of the 
transverse momentum $p_T$ for the inclusive production of $B^+$ 
mesons calculated in the GM-VFNS at $\sqrt{S} = 5.02$ TeV and 
$|y| < 2.4$ with the {\em original}€ scale choice $\mu_R = \mu_I 
= \mu_F = m_T$ (left panel) and with the {\em modified} scale choice 
$\mu_R = \mu_I = \mu_F = 0.5 m_T$ (right panel) compared to CMS 
data \cite{Khachatryan:2015uja}. 
}
\end{figure*}
%%%%%%%%%%%%%%%%%%%%%%%%%%%%%%%%%

\begin{figure*}
\begin{center}
\includegraphics[width=7.5cm]{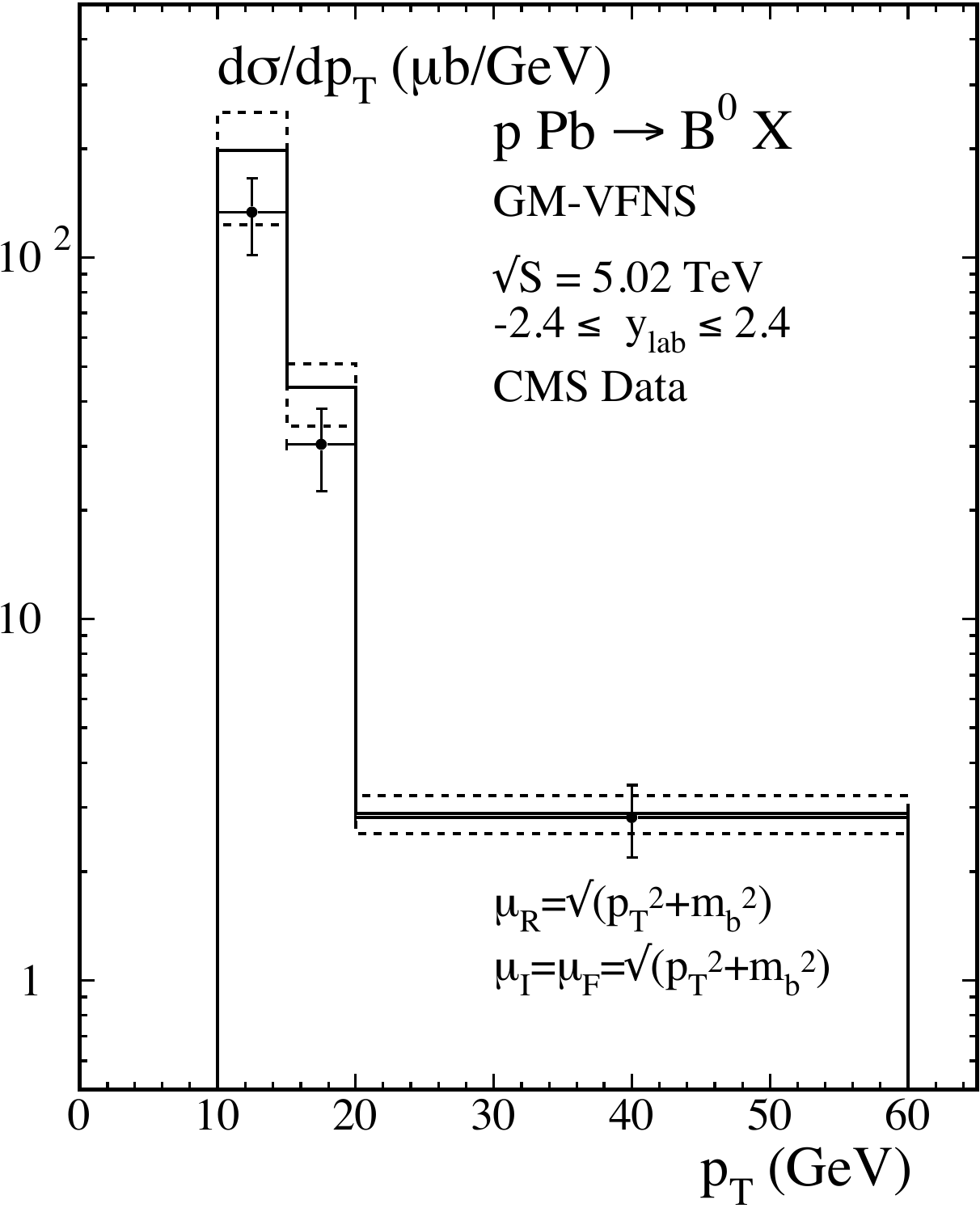}
\includegraphics[width=7.5cm]{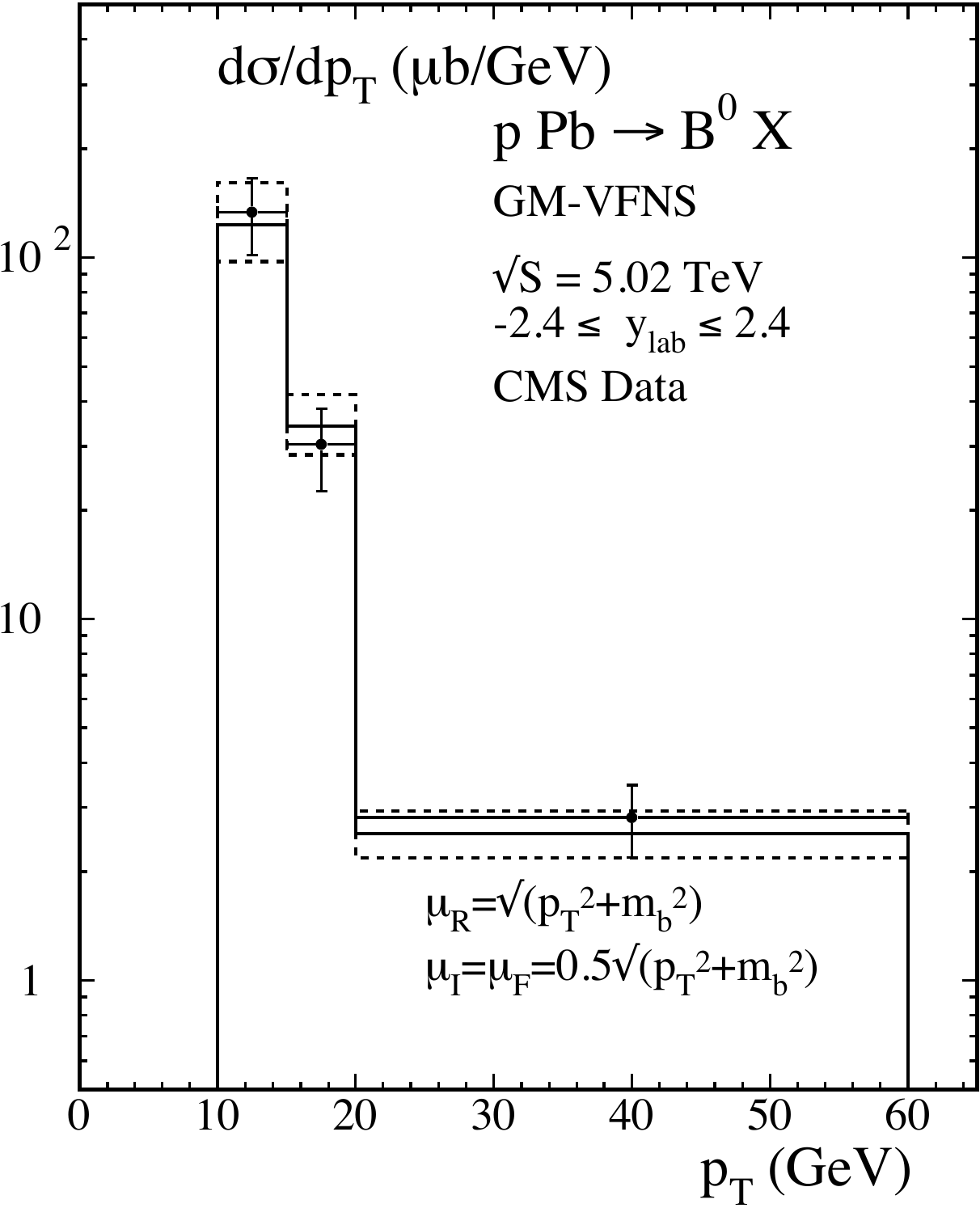}
\end{center}
\caption{ 
\label{fig:8} 
Differential cross section $A d\sigma/dp_T$ as a function of the 
transverse momentum $p_T$ for the inclusive production of $B^0$ 
mesons calculated in the GM-VFNS at $\sqrt{S} = 5.02$ TeV and 
$|y| < 2.4$ with the {\em original}€ scale choice $\mu_R = \mu_I 
= \mu_F = m_T$ (left panel) and with the {\em modified} scale choice 
$\mu_R = \mu_I = \mu_F = 0.5 m_T$ (right panel) compared to CMS 
data \cite{Khachatryan:2015uja}. 
}
\end{figure*}

\begin{figure*}
\begin{center}
\includegraphics[width=7.5cm]{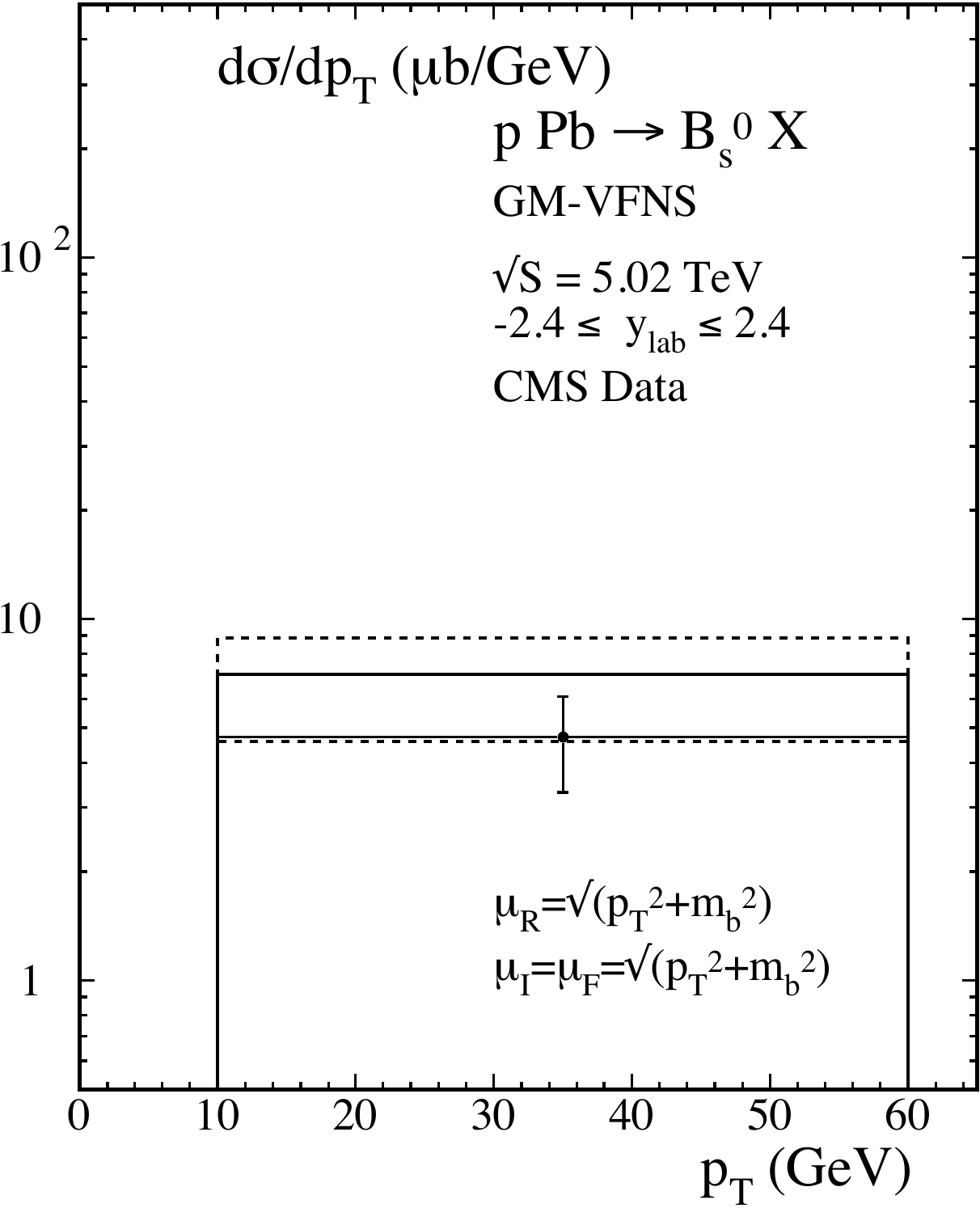}
\includegraphics[width=7.5cm]{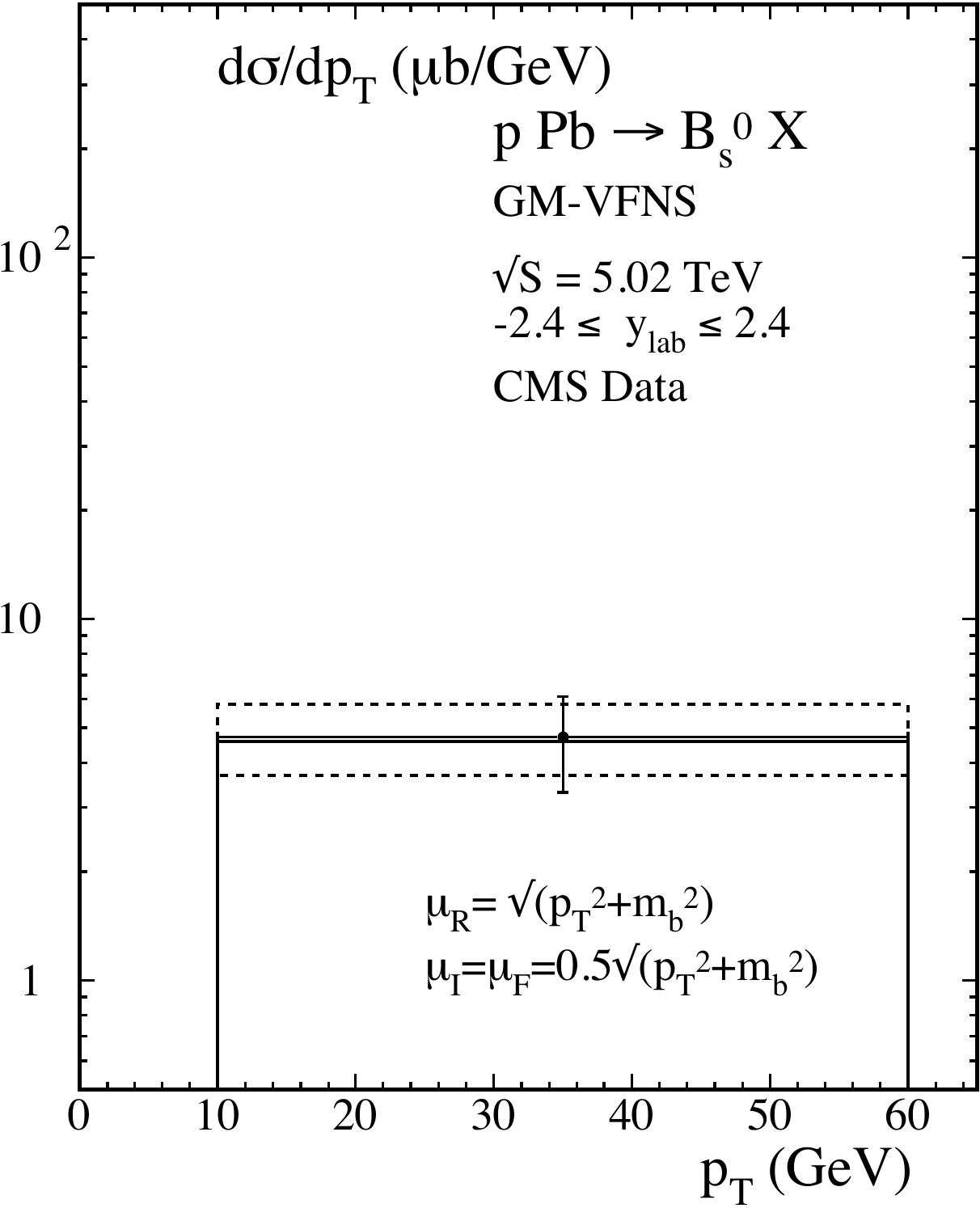}
\end{center}
\caption{ 
\label{fig:9} 
Differential cross section $A d\sigma/dp_T$ as a function of the 
transverse momentum $p_T$ for the inclusive production of $B_s^0$ 
mesons calculated in the GM-VFNS at $\sqrt{S} = 5.02$ TeV and 
$|y| < 2.4$ with the {\em original}€ scale choice $\mu_R = \mu_I 
= \mu_F = m_T$ (left panel) and with the {\em modified} scale choice 
$\mu_R = \mu_I = \mu_F = 0.5 m_T$ (right panel) compared to CMS 
data \cite{Khachatryan:2015uja}. 
} 
\end{figure*}

In the following we show the results for  $Ad\sigma/dp_T$ at 
$\sqrt{S} = 5.02$ TeV in the rapidity interval $|y| < 2.4$, 
again obtained from the p-p cross section $d\sigma/dp_T$ by 
multiplication with the mass number $A$. We have done these 
calculations for the {\em original}€ scale choice $\mu_o = 
\sqrt{m_b^2 + p_T^2}$; for the {\em modified} choice we 
decided to choose $\mu_m = 0.5 \sqrt{m_b^2 + p_T^2} = 0.5 
\mu_o$ in order to allow for a direct comparison with the 
previous work \cite{Kniehl:2015fla}\footnote{
  The value of $\mu_{I,F}$ at $p_T=0$ is not very relevant 
  here since we will compare with data at large $p_T$. With 
  $\mu_{I,F} = 0.5 \sqrt{4 m_b^2 + p_T^2}$ the cross section 
  would increase by only 12\,\% in the first $p_T$-bin 
  (10 GeV $ \leq p_T \leq 15$ GeV) and by less than 2\,\% at 
  higher $p_T$.}.  
$m_b$ is the bottom quark mass, $m_b = 4.5$ GeV. The FF for 
$b \to B$ was taken from \cite{Kniehl:2008zza} for all 
three $B$ meson states. Cross sections for the different $B$ 
meson species differ only by their respective constant 
fragmentation fractions. Our results are compared to the CMS 
data for p-Pb collisions \cite{Khachatryan:2015uja} and are 
shown for $B^+$, $B^0$ and $B_s^0$ production, respectively, 
in the left panels of Figs.\ \ref{fig:7}, \ref{fig:8}, and 
\ref{fig:9} for $\mu = \mu_o$ and in the right panels of these 
figures for $\mu = \mu_m$. As to be expected the results for 
the {\em original}€ scale choice $\mu_o$ lie slightly higher 
than for the {\em modified} scale choice $\mu_m$, but the 
difference is decreasing towards larger $p_T$. For all cases 
data and theory agrees within theoretical and experimental errors.

%%%%%%%%%%%%%%%%%%%%%%%%%%%%%%%%%
%%%%%%%%%%%%%%%%%%%%%%%%%%%%%%%%%

%%%%%%%%%%%%%%%%%%%%%%%%%%%%%%%%%
%%%%%%%%%%%%%%%%%%%%%%%%%%%%%%%%%

\begin{figure*}[b!]
\begin{center}
\includegraphics[width=7.5cm]{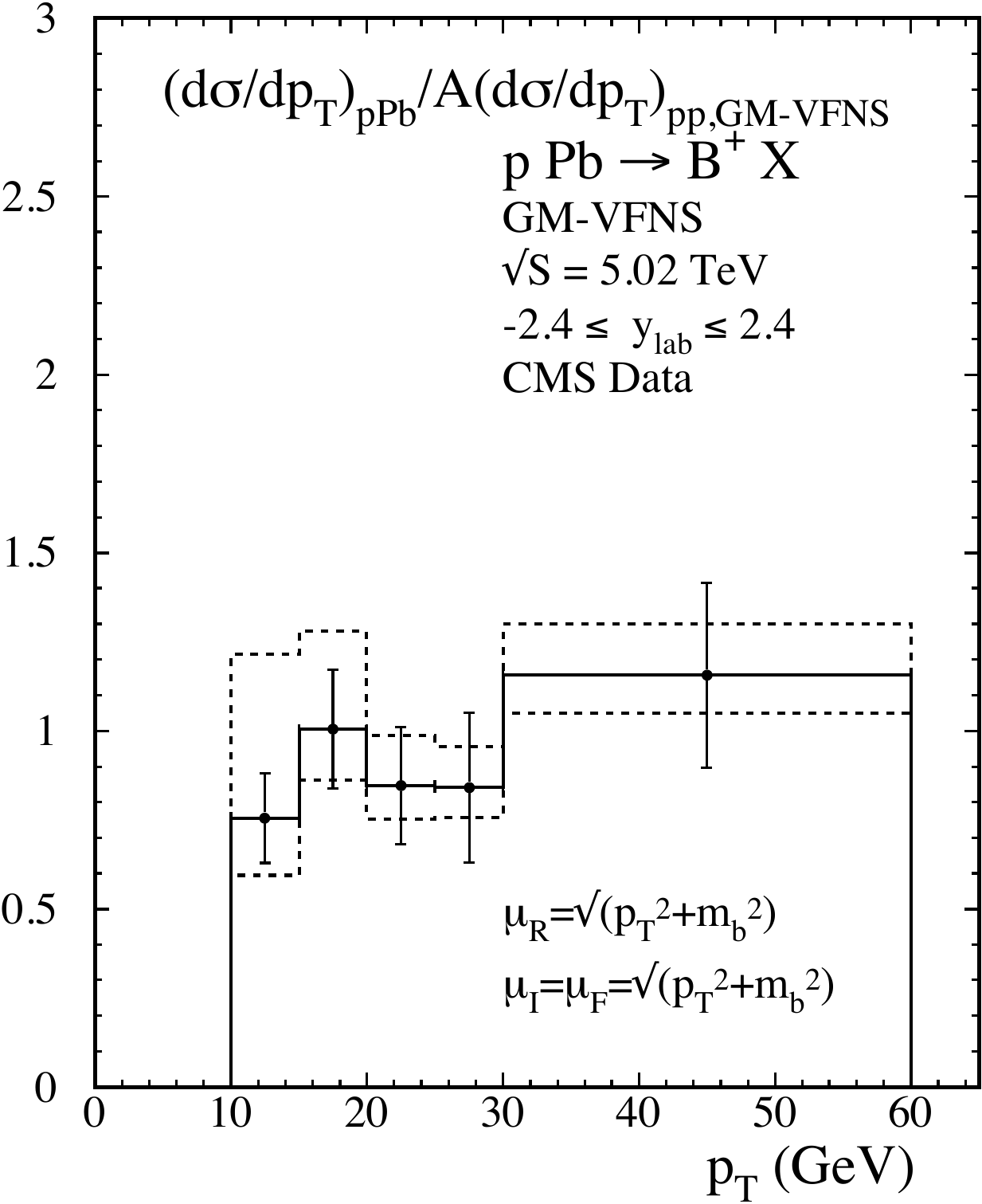}
\includegraphics[width=7.5cm]{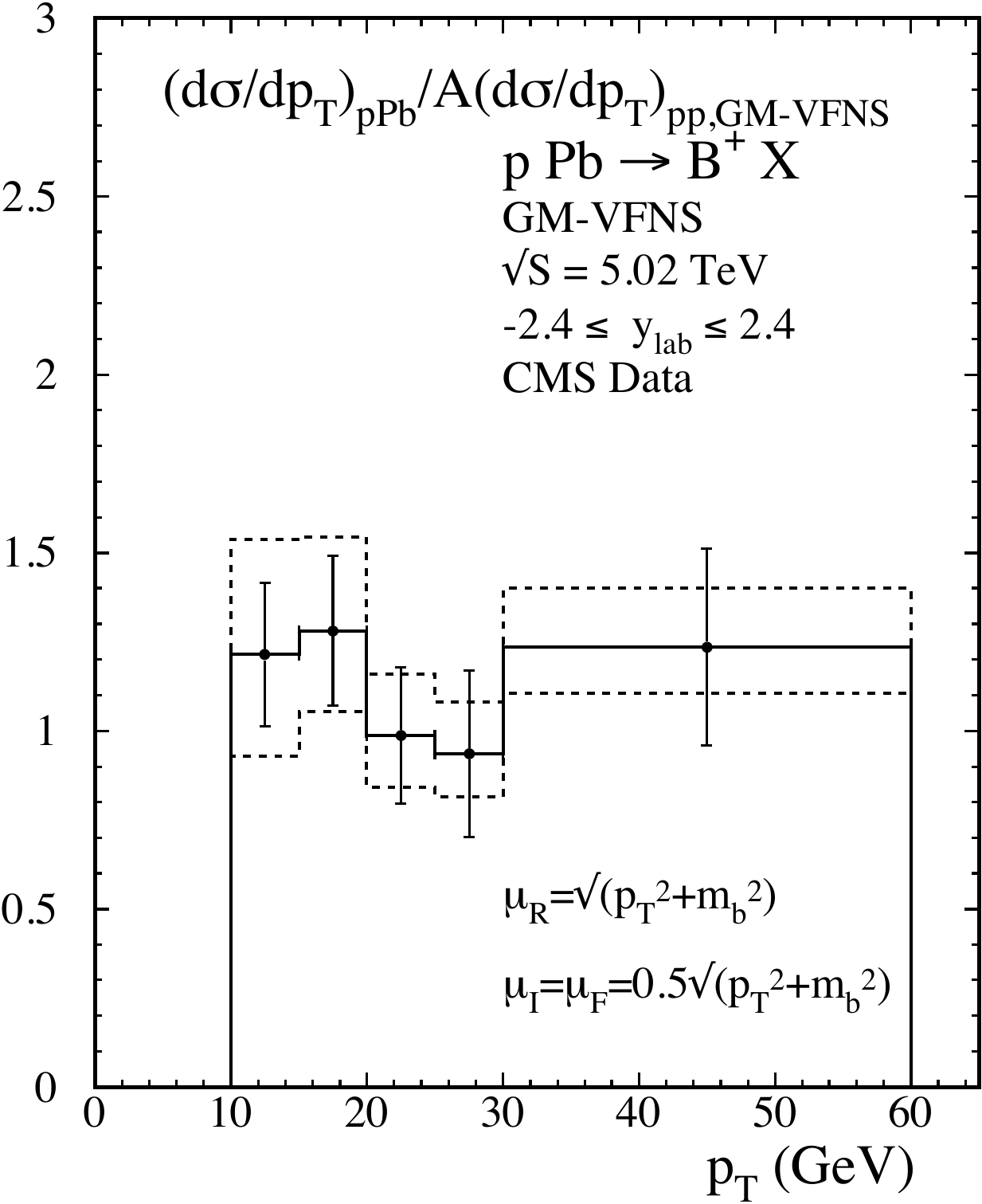}
\end{center}
\caption{ 
\label{fig:10} 
Ratio of the measured CMS cross section $d\sigma/dp_T$ to the 
GM-VFNS cross section shown in Fig.\ \ref{fig:7} for the 
{\em original}€ scale choice (left panel) and for the {\em modified} 
scale choice (right panel) for inclusive $B^+$ production. 
}
\end{figure*}

\begin{figure*}
\begin{center}
\includegraphics[width=7.5cm]{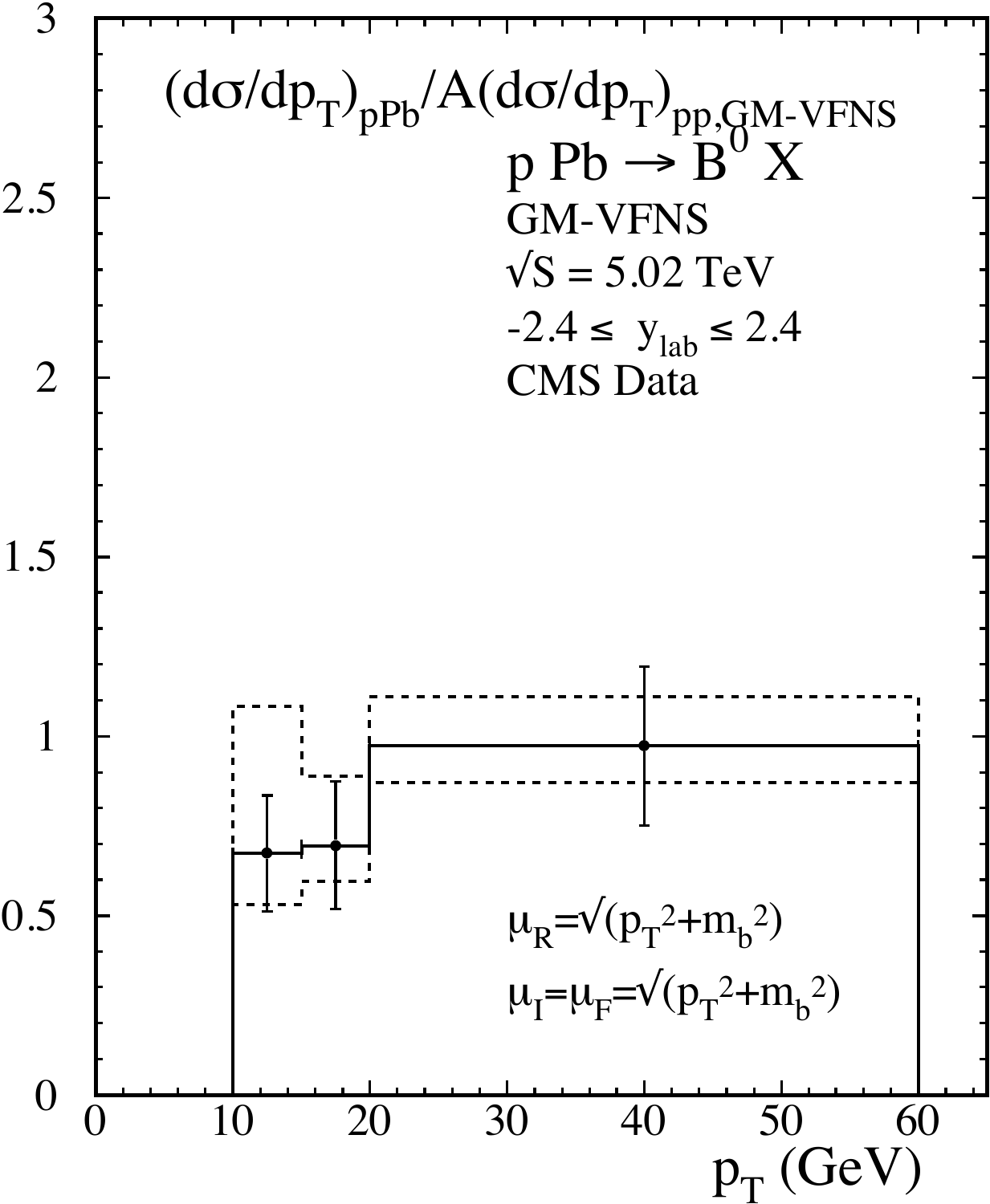}
\includegraphics[width=7.5cm]{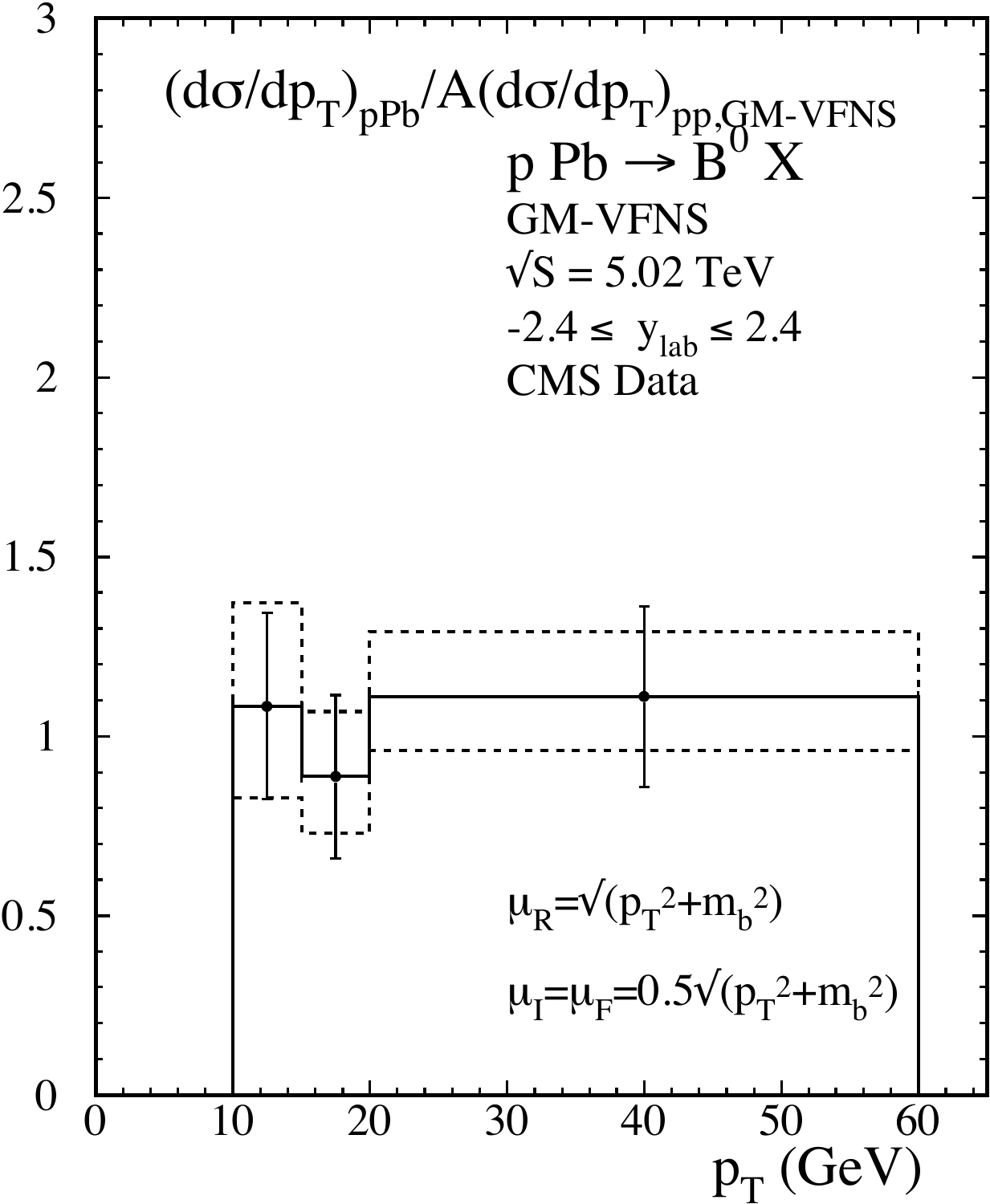}
\end{center}
\caption{
\label{fig:11} 
Ratio of the measured CMS cross section $d\sigma/dp_T$ to the 
GM-VFNS cross section shown in Fig.\ \ref{fig:8} for the 
{\em original}€ scale choice (left panel) and for the {\em modified} 
scale choice (right panel) for inclusive $B^0$ production. 
}
\end{figure*}

\begin{figure*}
\begin{center}
\includegraphics[width=7.5cm]{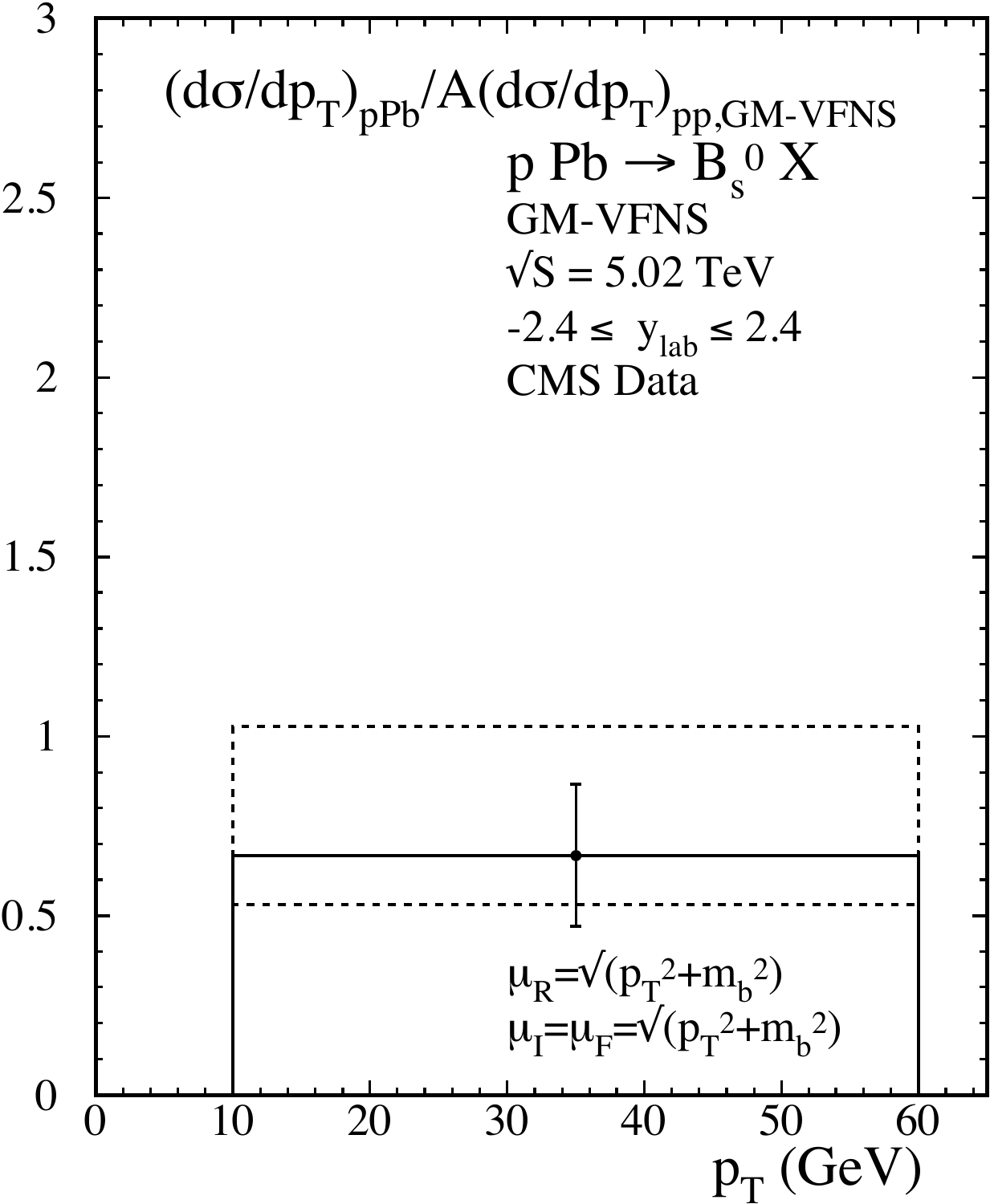}
\includegraphics[width=7.5cm]{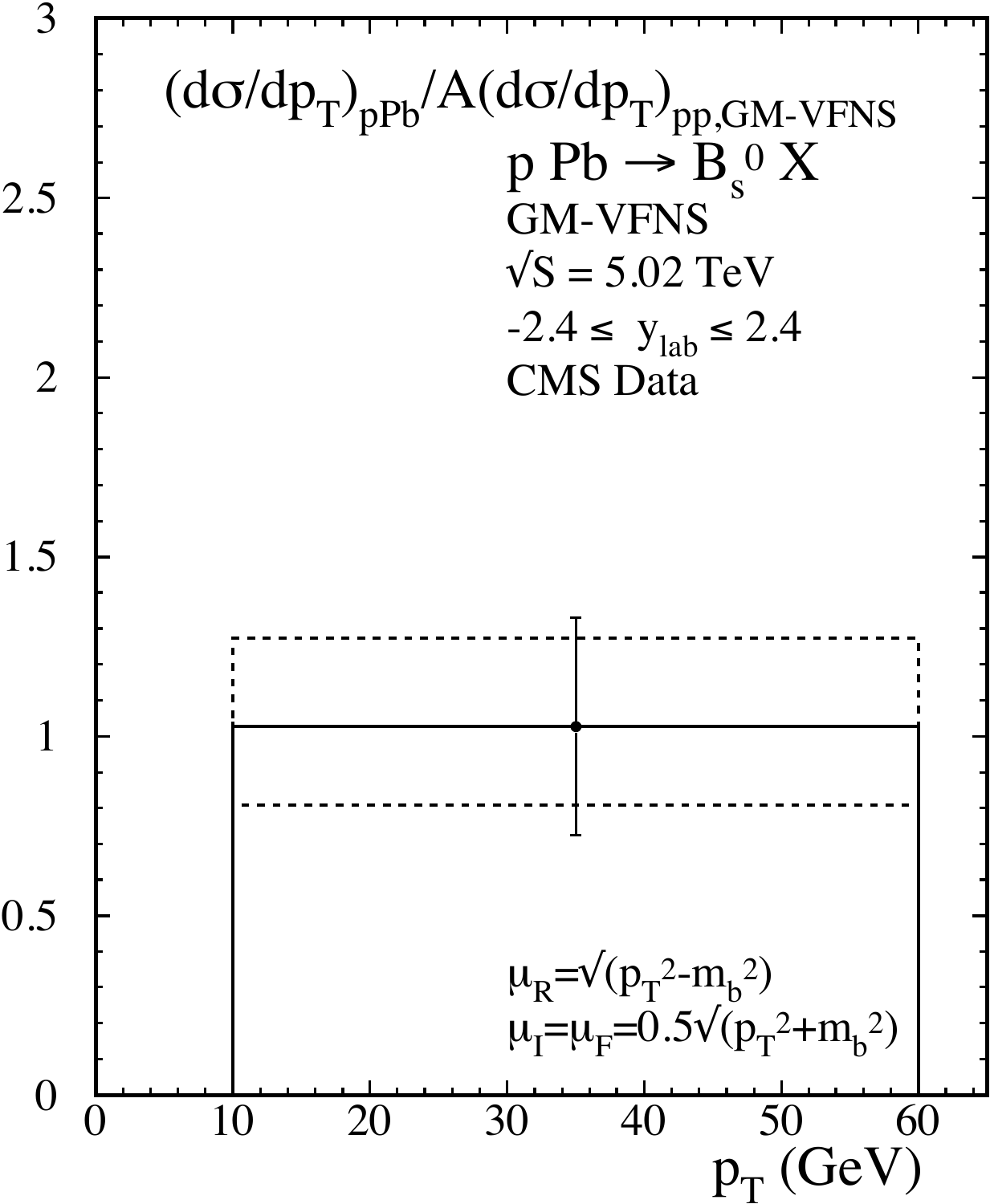}
\end{center}
\caption{
\label{fig:12} 
Ratio of the measured CMS cross section $d\sigma/dp_T$ to the 
GM-VFNS cross section shown in Fig.\ \ref{fig:9} for the 
{\em original}€ scale choice (left panel) and for the {\em modified} 
scale choice (right panel) for inclusive $B_s^0$ production. 
}
\end{figure*}

The comparison between the experimental cross section $d\sigma / 
dp_T$ for p-Pb scattering and the theoretical cross sections 
$Ad\sigma/dp_T$ becomes more clear when presented in terms of 
the nuclear modification factors $R_{\rm pPb} = 
(d\sigma/dp_T)_{\rm pPb} / A(d\sigma/dp_T)_{\rm pp}$. We show 
these ratios for all three $B$ meson species and for 
both scale choices, $\mu_o$ and $\mu_m$, in Figs.\ \ref{fig:10}, 
\ref{fig:11}, and \ref{fig:12} (left and right panels). We 
notice that with the {\em modified} scale choice, the ratio 
$R_{\rm pPb}$ agrees with one within experimental errors, 
even without taking into account the theory uncertainty due 
to scale variations given by the dashed lines in Figs.\ 
\ref{fig:10}-\ref{fig:12}. For the {\em modified} scale choice 
our results agree also rather well with those presented in 
\cite{Khachatryan:2015uja} where the p-p cross section used 
to obtain $R_{\rm pPb}$ was calculated in the FONLL approach 
\cite{Cacciari:2012ny}.

%%%%%%%%%%%%%%%%%%%%%%%%%%%%%%%%%
%%%%%%%%%%%%%%%%%%%%%%%%%%%%%%%%%

Our results for the nuclear modification factor $R_{\rm pPb}$ 
compared with CMS data differ somewhat for the two scale 
choices $\mu = \mu_o$ and $\mu = \mu_m$ (compare left and 
right panels of Figs.\ \ref{fig:10}, \ref{fig:11}, 
\ref{fig:12}). For $\mu = \mu_m$ the ratios $R_{\rm pPb}$ 
are equal to one for all bins within the precision of the 
data. For the {\em original}€ scale choice $\mu = \mu_o$ 
deviations from one seem to occur already within present 
errors in some of the low-$p_T$ bins (see left panels of 
Figs.\ \ref{fig:10}, \ref{fig:11}, and \ref{fig:12}). 
However, the observed deviations would become significant 
only if the experimental errors could be reduced, by at 
least a factor of two. It seems obvious to us that also theory 
uncertainties will have to be reduced before a conclusive 
interpretation of the data will be possible. This will 
require the calculation of higher-oder corrections which 
are expected to reduce the uncertainties due to the choice 
of renormalization and factorization scales. 

%%%%%%%%%%%%%%%%%%%%%%%%%%%%%%%%%
%%%%%%%%%%%%%%%%%%%%%%%%%%%%%%%%%

%%%%%%%%%%%%%%%%%%%%%%%%%%%%%%%%%
%%%%%%%%%%%%%%%%%%%%%%%%%%%%%%%%%

\clearpage

%%%%%%%%%%%%%%%%%%%%%%%%%%%%%%%%%%%%%%%%%%%%%%%%%%%%%%%%%%%%%%%%%%%%%%%
\section{Conclusions}

We have studied $D$ and $B$ meson production in p-Pb collisions 
and made, for the first time, a comparison with predictions 
obtained at NLO in the GM-VFNS. Our main results are shown in 
the right panels of Figs.\ \ref{fig:2} - \ref{fig:5}, \ref{fig:L6} 
for $D$-meson production and in Figs.\ \ref{fig:10} - \ref{fig:12} 
for $B$-meson production. The comparison with data confirms our 
previous findings that a suitable choice of the factorization 
scale parameters can be found which brings the experimental data 
obtained by the LHC collaborations ALICE, CMS and LHCb into good 
agreement with predictions obtained in the general-mass 
variable-flavour-number scheme. 

The ratio of data for p-Pb collisions over theory predictions
for $A$ times p-p cross sections is an important observable 
which could provide information about the nuclear modification 
of parton distribution functions, for example due to 
initial-state interaction effects. We found that for charmed 
meson production, the ratios of data over theory predictions 
at $p_T > 6$ GeV are compatible with one within uncertainties 
and deviations are not larger than 40\%. At small transverse 
momenta, $p_T < 6$ GeV, the ratios for data from ALICE at 
mid-rapidity increase to values of about 1.5 and larger. The 
data from the LHCb collaboration for $D$-production in the 
forward region, however, do not show such a strong enhancement 
of the nuclear modification ratio. Interestingsly, the 
ratio of p-p data over theory show deviations from one of 
the same size and with a similar $p_T$-dependence. It will 
be interesting to include forthcoming more precise data in 
our analysis, as for example from Ref.\ \cite{Acharya:2017jgo}.

Experimental uncertainties are often still large, but data 
are steadily improving. In particular the most recent data 
from LHCb are promising and one can expect that updated fits 
of nuclear PDFs with smaller uncertainties than the existing 
parametrizations will be possible. At present, however, scale 
uncertainties are still very large and it is therefore 
doubtful whether the observed deviations can be interpreted 
as due to nuclear modification effects. Higher precision of 
the measurements as well as of theory predictions is needed 
in order to draw firm conclusions.

%%%%%%%%%%%%%%%%%%%%%%%%%%%%%%%%%%%%%%%%%%%%%%%%%%%%%%%%%%%%%%%%%%%%%%%

\end{document}